\documentclass[showpacs,preprintnumbers,amsmath,amssymb,floatfix,nofootinbib]{revtex4}

\usepackage{color}   %option added by cyw eg \pagecolor{blue} \color{red}
\usepackage{graphicx}% Include figure files
\usepackage{dcolumn}% Align table columns on decimal point
\usepackage{bm}% bold math

\begin{document}

\def\bbox#1{\hbox{\boldmath${#1}$}}
\def\blambda{{\hbox{\boldmath $\lambda$}}}
\def\eeta{{\hbox{\boldmath $\eta$}}}
\def\bxi{{\hbox{\boldmath $\xi$}}}
\def\bzeta{{\hbox{\boldmath $\zeta$}}}

%\begin{CJK*}{GB}{gbsn} % Use default fonts from CJK (gbsn)

\title{ Momentum Kick Model Analysis of PHENIX Near-Side Ridge Data and
Photon Jet }

\author{Cheuk-Yin Wong\footnote{wongc@ornl.gov}
}

\affiliation{Physics Division, Oak Ridge National Laboratory, 
Oak Ridge, TN 37831}

\date{\today}

\begin{abstract}
  We analyze PHENIX near-side ridge data for central Au+Au collisions
  at $\sqrt{s_{NN}}=200$ GeV with the momentum kick model, in which a
  near-side jet emerges near the surface, kicks medium partons, loses
  energy, and fragments into the trigger particle and fragmentation
  products. The kicked medium partons subsequently materialize as the
  observed ridge particles, which carry direct information on the
  early parton momentum distribution and the magnitude of the momentum
  kick.  We find that the PHENIX ridge data can be described well by
  the momentum kick model and the extracted early partons momentum
  distribution has a thermal-like transverse distribution and a
  rapidity plateau structure.  We also find that the parton-parton
  scattering between the jet parton and the medium parton involves the
  exchange of a non-perturbative pomeron, for jet partons in momentum
  range considered in the near-side ridge measurements.

\end{abstract}


\maketitle

%\end{CJK*}

\section{Introduction}

Recently, the STAR Collaboration
\cite{Ada05,Ada06,Put07,Bie07,Wan07,Bie07a,Abe07,Mol07,Lon07,Nat08,Fen08,Net08,Bar08}
observed a $\Delta \phi$-$\Delta \eta$ correlation of particles
associated with a high-$p_t$ near-side hadron trigger particle in
central Au+Au collisions at $\sqrt{s_{NN}}=200$ GeV, where $\Delta
\phi$ and $\Delta \eta$ are the azimuthal angle and pseudorapidity
differences measured relative to the trigger particle,
respectively. Particles associated with the near-side jet can be
decomposed into a ``jet component'' at $(\Delta \phi, \Delta \eta)$$
\sim$(0,0), and a ``ridge component'' at $\Delta\phi$$ \sim$0 with a
ridge structure in $\Delta \eta$.  A similar correlation with a
high-$p_t$ trigger has also been observed by the PHENIX Collaboration
\cite{Ada08,Mcc08,Che08} and the PHOBOS Collaboration \cite{Wen08}.
Recent reviews of the ridge phenomenon have also been presented
\cite{Tan08,Jia08qm,Lee08}.

In this manuscript, we shall limit our attention to the ridge
phenomenon involving a high $p_t$ jet on the near-side.  We shall not
consider ridge-type $\Delta \phi$-$\Delta \eta$ correlations that have
also been observed between two low-$p_t$ hadrons \cite{Dau08}, as they
do not involve the occurrence of a high-$p_t$ jet on the near-side.

Many theoretical models
\cite{Won07,Won07a,Won08,Won08a,Won09,Shu07,Vol05,Chi08,Hwa03,Chi05,Hwa07,Pan07,Dum08,Gav08,Gav08a,Arm04,Rom07,Maj07,Dum07,Miz08,Jia08,Jia08a}
have been proposed to discuss the ridge phenomenon.  The model of
Ref.\ \cite{Shu07} assumes that the ridge particles arise from the
extra particles deposited by the forward and backward beam jets at the
source point associated with the two transverse jets.  The correlation
of the jet source transverse position and the transverse medium flow
then leads to an azimuthal distribution with a width in $\Delta \phi$
\cite{Vol05,Shu07}.  The width in $\Delta \phi$ obtained from such a
model is wide in comparison with experimental data \cite{Shu07}.  The
Correlated Emission Model \cite{Hwa03,Chi05,Hwa07,Chi08} assumes that
ridge particles arise from soft thermal gluons radiated along the jet
direction, with an enhancement due to the radial flow.  The models of
Refs.\ \cite{Shu07} and \cite{Chi08} deal with the azimuthal
correlations in the central rapidity region, and the pseudorapidity
correlation has not yet been considered.  The back-splash model
assumes that the ridge on the near-side arises from the hydrodynamical
back-splash of the away-side jet flow \cite{Pan07}; hydrodynamical
calculations for such a model has not yet been made. The Glasma model
examines $\Delta \phi$-$\Delta \eta$ correlation between two low-$p_t$
hadrons \cite{Dau08} without a high-$p_t$ trigger and assumes that the
ridge in soft low-$p_t$ pairs arises from the initially
boost-invariant distribution that persists in the bulk matter for
low-$p_t$ particles \cite{Dum08,Gav08,Gav08a}.  The Jet Broadening
models \cite{Rom07,Arm04,Maj07,Dum07} consider the ridge particles as
arising from radiated gluons of the incident jet; they have not been
compared quantitatively with the ridge data.  Taking the features of
jet broadening as free parameters in a hydrodynamical calculation
leads to a theoretical jet peak to ridge ratio large in comparison
with experiment \cite{Miz08}.  The possibility of the intermediate
$p_t$ trigger arising from the medium-medium recombination adds further
complications to the analysis of the ridge phenomenon
\cite{Jia08,Jia08a}.  Recent PHOBOS observation that the ridge extends
to pseudorapidity separations as large as $|\Delta \eta|\!\sim$4
\cite{Wen08} provides an important test for the models.

Successful analyses of experimental near-side data have been obtained
in the momentum kick model, over large phase space of the associated
particles in $p_t$, $\Delta \phi,$ and $\Delta \eta$
\cite{Won07,Won07a,Won08,Won08a}.  In this model, the ridge particles
are described as arising from partons in the medium that are kicked by
the jet.  We envisage that a near-side jet emerges near the surface,
kicks (or scatters) medium partons, loses energy, and fragments into
the trigger particle and fragmentation products.  By assumption of
parton-hadron duality, the kicked (or scattered) medium partons
subsequently materialize as the observed ridge particles, which can be
used to extract valuable information on the jet-medium interaction and
the properties of the early parton medium.  In the description of the
interaction between the medium and a jet in the momentum kick model,
we have chosen to represent the medium as particles instead of fields,
because of (i) the short-range nature of the color screening
interaction between partons in a dense color medium \cite{Kac05,Won02}
and (ii) the observed azimuthal kinematic correlation between the
ridge particles and the trigger jet.

Our task can be made easier here as we can divide the theoretical
analysis in three steps.  The first step is to set up the basic
phenomenological theory of the momentum kick model in which physical
quantities enter as important parameters.  The second step consists of
comparing the extracted physical quantities with those in other
observed phenomena.  The third step consists of constructing
fundamental theoretical models that can explain these physical
quantities.

Following such a strategy, we describe the production process of
associated particles as consisting of the jet component and the ridge
component.  The ridge component depends on the magnitude of the
momentum kick, the number of jet-(medium parton) collisions, and the
shape of the early medium parton momentum distribution.  On the other
hand, the jet component yield per trigger in a nucleus-nucleus
collision can be described as an attenuated jet component of a $pp$
collision.  It is therefore necessary to analyze the auxiliary
associated particle yield in $pp$ collisions in order to specify the
jet component in nucleus-nucleus collisions.

Our successful description of the experimental data allows us to
extract physical quantities from STAR near-side ridge data in central
Au+Au collisions at $\sqrt{s_{NN}}=200$ GeV \cite{Won08a}.  In the
process, we infer that the shape of the early parton momentum
distribution possesses a thermal-like transverse distribution and a
rapidity plateau structure.  We find that the magnitude of the
longitudinal momentum kick is about 1 GeV per jet-(medium parton)
collision.  We infer also that for a central Au+Au collision the
number of jet-(medium parton) collision multiplied by the attenuation
factor is about 4.  As not much is known about these physical
quantities, the extracted quantities provide useful insight into the
properties of the early partons and their interactions with the jet in
nucleus-nucleus collisions.

With the successes in analyzing the STAR near-side ridge data, it is
of interest to see whether the momentum kick model is consistent with
other experimental measurements.  Our first test of the momentum kick
model gives a good prediction \cite{Won08,Won08a} of the PHOBOS data
\cite{Wen08} at large rapidities, indicating the approximate validity
of the momentum kick model and the presence of the rapidity plateau.

We wish to analyze here the PHENIX ridge data which cover a smaller
region of pseudorapidities, $|\eta^{\rm trig},\eta^{\rm assoc}|<$0.35,
but a large number of $p_t^{\rm trig} \otimes p_t^{\rm assoc}$
combinations. Both the jet and ridge components contribute and
interplay in the small $\Delta \eta$ region on the near-side of the
jet.  This is different from the STAR and PHOBOS ridge data which
cover a large range of pseudorapidities.  The jet component is
important at $\Delta \eta \sim 0$ whereas the ridge component
dominates for $|\Delta \eta| >0.7$.

After extracting the physical quantities from the analysis of the
PHENIX ridge data, we wish to find out the nature of the scattering
between the jet parton and the medium parton.  In the experimental set
up, jet triggers have been accepted in the interval $2< p_t^{\rm trig}
< 10$ GeV. The incident jet parton has an initial transverse momentum
of order $p_t^{\rm jet}\sim$10 GeV, as a jet parton loses a few GeV in
kicking a few medium partons.  Is the scattering between a jet parton
and a medium parton a perturbative or non-perturbative QCD scattering
process, for jet partons in this momentum range?  Our ability to
ascertain the nature of the parton-parton scattering will help us
select the proper description to formulate the process of energy loss
for these jet partons.

Previously, phenomenological model of hadron-hadron differential cross
section in terms of parton-parton collisions with a finite correlation
length was successfully applied in the modified Chou-Yang model
\cite{Sch81,Cho68,Bia76,Lev74}.  In recent years, much progress has
been made on the description of non-perturbative parton-parton
scattering, in connection with a better understanding on the nature of
the non-perturbative soft pomeron
\cite{Lan88,Nac91,Dos87,Kra90,Dos92,Dos01,Ber99,DiG92,Meg99,For97,Don02}.
In particular, hadron-hadron elastic differential cross section
analysis and lattice gauge calculations support the concept of the
structure of a pomeron with a small correlation length.  These recent
theoretical advances will allow us to compare the characteristics of
the parton-parton scattering in the present momentum kick model with
those parton-parton collisions arising from the exchange of
non-perturbative pomerons.

Turning to the properties of the early parton momentum distribution
extracted from the momentum kick model, we note that the presence of
the rapidity plateau in the early history of a central nucleus-nucleus
collision as inferred from the momentum kick model is not a surprising
result, as the rapidity plateau structure occurs in elementary
processes involving the fragmentation of flux tubes
\cite{Cas74,Bjo83,Won91,Won94,Won09} and in many particle production
models such as models based on preconfinement \cite{Wol80},
parton-hadron duality \cite{Van88} cluster fragmentation \cite{Odo80},
string-fragmentation \cite{And83}, dual-partons \cite{Cap78}, the
Venus model \cite{Wer88}, the RQMD model \cite{Sor89}, multiple
collision model \cite{Won89}, parton cascade model \cite{Wan94,Gei92},
color-glass condensate model \cite{McL94}, the AMPT model \cite{Li96},
the Lexus model \cite{Jeo97}, and many others.  To investigate the
origin of the rapidity plateau in a quantum mechanical framework, we
can go a step further to use the physical argument of transverse
confinement to establish a connection between QCD and QED2 (Quantum
Electrodynamics in 2-dimensions) \cite{Won08a,Won09}.  One finds that
a rapidity plateau of produced particles is a natural occurrence when
color charges pull away from each other at high energies
\cite{Cas74,Bjo83,Won91,Won94,Won09} as in QED2
\cite{Sch62,Low71,Col75,Col76}.  Experimental evidence for a rapidity
plateau along the sphericity axis or the thrust axis has been observed
earlier in $\pi^\pm$ production in high-energy $e^+$-$e^-$
annihilations \cite{Aih88,Hof88,Pet88,Abe99,Abr99}.  A rapidity plateau
structure has also been observed in $pp$ collisions at RHIC energies
by the BRAHMS Collaboration \cite{Yan08}.

In addition to the magnitude of the momentum kick acquired by a medium
parton per jet-(medium parton) collision, the ridge data also give
information on the number of kicked partons.  These physical
quantities are clearly related to the energy loss of a jet in the
dense medium.  A consistent picture of both the ridge yield and jet
quenching emerges from the momentum kick model analyses \cite{Won08a}
and complements other studies of the jet quenching phenomenon
\cite{Jetxxx}.

The analysis of the PHENIX near-side ridge data also provides an
opportunity to examine an additional test of the momentum kick model
using a high-$p_t$ photon jet. Nucleon-nucleon collisions can lead to
the occurrence of a high-$p_t$ parton jet in coincidence with a
high-$p_t$ photon jet.  In a central nucleus-nucleus collision with an
away-side parton jet, we can use a photon jet on the near-side to test
different ridge models \cite{Cia08}.  By comparing associated
particles on the near-side with a hadron or a photon jet, we can
separate out effects owing to the collision of the near-side jet.  If
the ridge arises from the medium as a result of the collision of the
near-side parton jet, as in the momentum kick model, the substitution
of a near-side photon jet will lead to a greatly-reduced yield of
ridge particles.  If the ridge particles arise from ``several extra
particles deposited by forward-backward beam jets into the fireball'',
as in the position-flow model of Ref.\ \cite{Shu07} and \cite{Vol05}
then the near-side ridge will remain for a near-side photon jet. If
the ridge arises from the back splash of the propagation of away-side
parton jet, then the large ridge structure will remain for a near-side
photon jet.  It is therefore of interest to make theoretical estimates
of the ridge yield in association with a near-side, high-$p_t$ photon
jet in the momentum kick model so as to assist experiments in such an
analysis.

This paper is organized as follows.  In Section II, we review and
summarize the main results of the momentum kick model.  In Section
III, we discuss the jet component in Au+Au and in $pp$ collisions.
The auxiliary associated particles yield in $pp$ collisions is
parametrized to assist the analysis of the ridge component in Au+Au
collisions.  In Section IV, momentum kick model description of the
ridge yield is presented and physical parameters are introduced to
describe the ridge component.  In Section V, we compare theoretical
and experimental results of the total associated particle yield in
PHENIX experiments using hadron triggers in different $p_t^{\rm trig}$
intervals.  In Section VI, VII, and VIII, we examine new insights
derived from the physical quantities extracted from the momentum kick
model.  Specifically, in Section VI, we find that the magnitude of the
longitudinal momentum kick along the jet direction $q_L$ is consistent
with the characterization that the scattering between the jet parton
and the medium parton involves the exchange of a non-perturbative
pomeron.  In Section VII, we find that the extracted shape of the
rapidity plateau of the early parton distribution is in between those
of the pp and Au+Au collisions, indicating an intermediate stage of
parton evolution.  In Section VIII, we find that the number of kicked
partons at the most central collision can provide the correct
normalization for the momentum kick model to describe the centrality
dependence of the ridge yield.  These results supports the approximate
validity of the momentum-kick model.  In Section IX, we calculate the
ridge yield when a high-$p_t$ photon jet occurs.  The results can be
used to guide our search for ways to discriminate different models.
In Section X, we present our discussions and conclusions.

\section{Review of the  Momentum Kick Model}

We shall briefly summarize the basic concepts of the momentum kick
model.  In the phenomenon of the ridge associated with the near-side
jet, it is observed that (i) the ridge particle yield increases with
the number of participants, (ii) the ridge yield appears to be nearly
independent of the trigger jet properties, (iii) the baryon to meson
ratios of the ridge particles are more similar to those of the bulk
matter than those of the jet, and (iv) the slope parameter of the
transverse distribution of ridge particles is intermediate between
those of the jet and the bulk matter \cite{Ada05}-\cite{Lee08}. These
features suggest that the ridge particles are medium partons, at an
early stage of the medium evolution during the passage of the jet.
The azimuthal correlation of the ridge particle with the jet and the
presence of strong screening suggest that the associated ridge
particle and the trigger jet are related by collisions.  A momentum
kick model was put forth to explain the ridge phenomenon
\cite{Won07,Won07a,Won08,Won08a,Won09}.

The model assumes that a near-side jet occurs near the surface,
collides with medium partons, loses energy along its way, and
fragments into the trigger and its associated fragmentation products
(the ``jet component'').  Those medium partons that collide with the
jet acquire a momentum kick along the jet direction.  They
subsequently materialize by parton-hadron duality as ridge particles
in the ``ridge component''.  In other words, the ridge particles are
medium partons kicked by the jet and they carry direct information on
the early parton momentum distribution and the magnitude of the
longitudinal momentum kick.

As described in detail in \cite{Won07,Won07a,Won08,Won08a}, we follow
a jet as it collides with medium partons in a dense medium and study
the yield of associated particles for a given $p_t^{\rm trig}$. The
evaluation of the ridge yield and the investigation of the quenching
of the jet will be greatly simplified by using average values of
various physical quantities, whose ``average'' attribute will be made
implicit.  

We label the normalized initial momentum distribution of medium
partons at the moment of jet-(medium parton) collisions by $E_i
dF/d{\bf p}_i$.  The jet imparts a momentum ${\bf q}$ onto a kicked
medium parton, which changes its momentum from ${\bf p}_i$ to ${\bf
  p}=(p_t,\eta,\phi)={\bf p}_f={\bf p}_i+{\bf q}$, as a result of the
jet-(medium parton) collision.  By assumption of parton-hadron
duality, the kicked medium partons subsequently materialize as
observed associated ridge particles.

We shall use the label ${\bf p}$ of the kicked medium partons
interchangeably with the label ${\bf p}^{\rm assoc}$ of associated
ridge particles.  The normalized final parton momentum distribution $E
dF/d{\bf p}$ at ${\bf p}$ is related to the normalized initial parton
momentum distribution $E_i dF/d{\bf p}_i$ at ${\bf p}_i$ at a shifted
momentum, ${\bf p}_i={\bf p}-{\bf q}$, and we have \cite{Won07}
\begin{eqnarray}
\label{final}
\frac{dF}{ p_{t}dp_{t}d\eta  d\phi} 
&=&\left [ \frac{dF}{ p_{ti}dp_{ti} dy_i d\phi_i }
  \frac{E}{E_i} \right ]_{{\bf p}_i ={\bf p}-{\bf q}}
  \sqrt{1-\frac{m^2}{(m^2+p_t^2) \cosh^2 y}},
\end{eqnarray}
where the factor $E/E_i$ ensures conservation of particle numbers and
the last factor changes the rapidity distribution of the kicked partons
to the pseudorapidity distribution \cite{Won94}.  

We characterize the number of partons kicked by the jet by
$\langle N_k \rangle$, which depends on the centrality and the
jet-(medium parton) cross section.  The (charged) ridge particle
momentum distribution in a central A+A collision per trigger is then
\begin{eqnarray}
\label{eq2}
\left [
\frac{dN_{\rm ch}}{ N_{\rm trig} p_t dp_t d\Delta \eta\,d\Delta \phi } 
\right ]_{\rm ridge}^{\rm AA}
&=&
\left [  f_R   \frac {2}{3} 
\langle N_k \rangle \frac { dF } {p_t dp_t\,
d\Delta \eta\, d\Delta \phi} \right ]_{\rm ridge}^{\rm AA} 
\nonumber \\ 
&=&  f_R   \frac {2}{3} 
\langle N_k \rangle 
\left [ \frac{dF}{ p_{ti}dp_{ti} dy_i d\phi_i } 
     \frac{E}{E_i} \right ]_{{\bf p}_i= {\bf p}-{\bf q}} 
\sqrt{1-\frac{m^2}{(m^2+p_t^2) \cosh^2 y}},
\end{eqnarray}
where $\Delta \eta=\eta-\eta^{\rm trig}$, $\Delta \phi=\phi-\phi^{\rm
  trig}$, $ f_R $ is the average survival factor for produced ridge
particles to reach the detector, and the factor $2/3$ is to indicate
that 2/3 of the produced associated particles (presumably pions) are
charged.\footnote {The charge fraction (2/3) assumed for a pion system
  can be modified for a medium with a more general composition.}
Present measurements furnish information only on the product
$f_R\langle N_k \rangle$.  The momentum kick ${\bf q}$ will be
distributed in the form of a cone around the trigger jet direction
with an average $\langle {\bf q}\rangle = q_L {\bf e}^{\rm trig}$
directed along the trigger direction ${\bf e}^{\rm trig}$,
characterized by the momentum kick magnitude $q_L$.  For brevity of
nomenclature, `the longitudinal momentum kick $q_L$ along the jet
direction' will henceforth be abbreviatingly called `the momentum kick
$q_L$'.

\section{The Jet Component in Au+Au and $pp$ collisions}

Experimental measurements of the associated particles in A+A
collisions include contributions from both the jet component and the
ridge component.  By comparing the associated particle yield per
trigger in central Au+Au collisions with the $pp$ associated particle
yield at $\Delta \eta\sim 0$, one finds that in the region of $p_t <
4$ GeV, the jet component yield in central Au+Au collisions can be
consistently described as an attenuated yield of associated particles
in a $pp$ collision \cite{Won08a},
\begin{eqnarray}
\label{obser}
\left [ 
\frac{1}{N_{\rm trig}}
\frac{dN_{\rm ch}} 
{p_{t} dp_{t} d\Delta \eta  d\Delta \phi } \right ]_{\rm jet}^{\rm AA} 
= 
 f_J  \frac { dN_{\rm jet}^{pp}} {p_t dp_t\, d\Delta \eta\, d\Delta
\phi} .
\end{eqnarray}
The survival factor $f_J$ varies with  $p_t^{\rm assoc}$ of the
associated particles, being relatively constant for low $p_t^{\rm
assoc}$ with a semi-empirical value of $f_J=0.632$ \cite{Won08a}.  It
reaches the value of unity when $p_t^{\rm assoc}$ of the associated
particle approaches $p_t^{\rm trig}$, corresponding to fragmentation
outside the medium (see the dependence of $f_J$ on $p_t^{\rm assoc}$
in Eq.\ (\ref{fj}) below).

To obtain the jet component in Au+Au collisions, we need the yield of
associated particles in a $pp$ collision.  In principle, the yield of
associated particles can be obtained from the description of jets in
perturbative QCD such as the Pythia computer program
\cite{Sjo01,Jia09a}.  The application of such a treatment with
different available sets of tuned parameters does not automatically
yield a perfect agreement of the theory with experiment.  Additional
fine tuning of many Pythia parameters and theoretical options is
needed \cite{Jia09a}.  Even with the fine tuning, the agreement of
theoretical results with both the experimental jet spectra and the
experimental associated particle correlation cannot be obtained
simultaneously at the present time \cite{Jia09a}.

For our purposes of studying the ridge phenomenon, the $pp$ associate
particle data are only auxiliary quantities that are needed to
calculate the total associated particle yield. One could in principle
make use of the experimental $pp$ data to infer the jet component of
the Au+Au jet component, with the help of Eq.\ (\ref{obser}). We shall
alternatively represent the experimental $pp$ data by simple
parametrization, which is just a short-hand way to stand for the
experimental $pp$ associated particle data, for the purpose of
assisting later the evaluation of the total associated particle yield.

The experimental associated particle distribution in $pp$ collisions
can be described well by \cite{Won08a}
\begin{eqnarray}
\label{jetfun}
\frac { dN_{\rm jet}^{pp}} {p_t dp_t\, d\Delta \eta\, d\Delta \phi}
\!\!= N_{\rm jet}
\frac{\exp\{(m-\sqrt{m^2+p_t^2})/T_{\rm jet}\}} {T_{\rm jet}(m+T_{\rm jet})}
\frac{1}{2\pi\sigma_\phi^2}
e^{- {[(\Delta \phi)^2+(\Delta \eta)^2]}/{2\sigma_\phi^2} },
\end{eqnarray}
where by assumption of hadron-parton duality $m$ is taken as the pion
mass $m_\pi$, $N_{\rm jet}$ is the total number of near-side (charged)
associated particles in a $pp$ collision, and $T_{\rm jet}$ is the jet
inverse slope (``temperature'') parameter of the ``$pp$ jet
component''.  In our search for parameter values we find that the
parameters $N_{\rm jet}$ and $T_{\rm jet}$ vary linearly with
$p_t^{\rm trig}$ of the trigger particle which we describe
as\footnote{In calculating the $N_{\rm jet}$ and $T_{\rm jet}$
  parameters using Eqs.\ (\ref{Njetv}) and (\ref{Tjetv}) for the
  interval of $p_t^{\rm trig}=$5-10 GeV, we use $\langle p_t^{\rm
    trig} \rangle =5.5$ GeV for this interval as the spectra of
  trigger particles decrease rapidly with $p_t^{\rm trig}$ and the
  dominant contributions for this interval come between $p_t^{\rm
    trig}=$5 and 6 GeV.}
\begin{eqnarray}
\label{Njetv}
N_{\rm jet} = N_{{\rm jet}0} + d_N ~ p_t^{\rm trig},
\end{eqnarray}
\begin{eqnarray}
\label{Tjetv}
T_{\rm jet}= T_{{\rm jet}0} + d_T ~ p_t^{\rm trig}.
\end{eqnarray}
We also find that the width parameter $\sigma_\phi$ depends slightly
on $p_t$ which we parametrize as
\begin{eqnarray}
\label{ma}
\sigma_\phi=\sigma_{\phi 0} \frac{m_a}{\sqrt{m_a^2+p_t^2}}.
\end{eqnarray}
We summarize the meaning of the parameters introduced to describe the
$pp$ associated particle data in Table I.
\vspace*{0.3cm} 
\begin{table}[h]
  \caption { Physical parameters in Eq.\ (\ref{jetfun}),
    for the description of associated particles in $pp$ collisions, and the meaning of each parameter }
\vspace*{0.3cm}
\begin{tabular}{|c|c|c|}
  \hline Category & Physical Parameter & Meaning 
  \\ \hline \hline   & $N_{\rm jet}$ & 
  number of associated particles per trigger in a $pp$ collision
  \\ \cline{2-3} 
  Properties of jet & $T_{\rm jet}$ & ``temperature'' of $p_t^{\rm assoc}$ distribution in a $pp$ collision
  \\ \cline{2-3} 
  particles associated & $\sigma_{\rm  \phi 0}$ & jet cone width parameter 
  \\ \cline{2-3} 
  with a trigger & $m_a$ & mass parameter to modify the variation of 
  \\ 
  in a $pp$ collision & & jet cone width $\sigma_{\rm \phi}$ with $p_t^{\rm assoc}$ 
  \\ \hline 
\end{tabular}
\end{table}
\vspace*{0.3cm} Using this set of parameters, we fit the $pp$
associated particle data obtained in PHENIX measurements for $pp$
collisions at $\sqrt{s_{NN}}=200$ GeV.  The values of the parameters
are given in Table II.  The theoretical results of $dN_{\rm
  ch}^{pp}/N_{\rm trig} d\Delta\phi$ are given as dash-dot curves in
Fig. 1 and the corresponding experimental data are represented by open
circles.  As one observes in Fig.\ 1, although the fit is not perfect,
the set parameters in Table II adequately describe the set of $pp$
associated particle data for $2< p_t^{\rm trig} < 10$ GeV and for $0.4 <
p_t^{\rm assoc}<4$ GeV.  The parametrization can be used to generate
the jet component for nucleus-nucleus collisions by assuming that the
jet component yield per trigger in a nucleus-nucleus collision is an
attenuated yield of the corresponding $pp$ collision.

\begin{figure} [h]
\includegraphics[angle=0,scale=0.70]{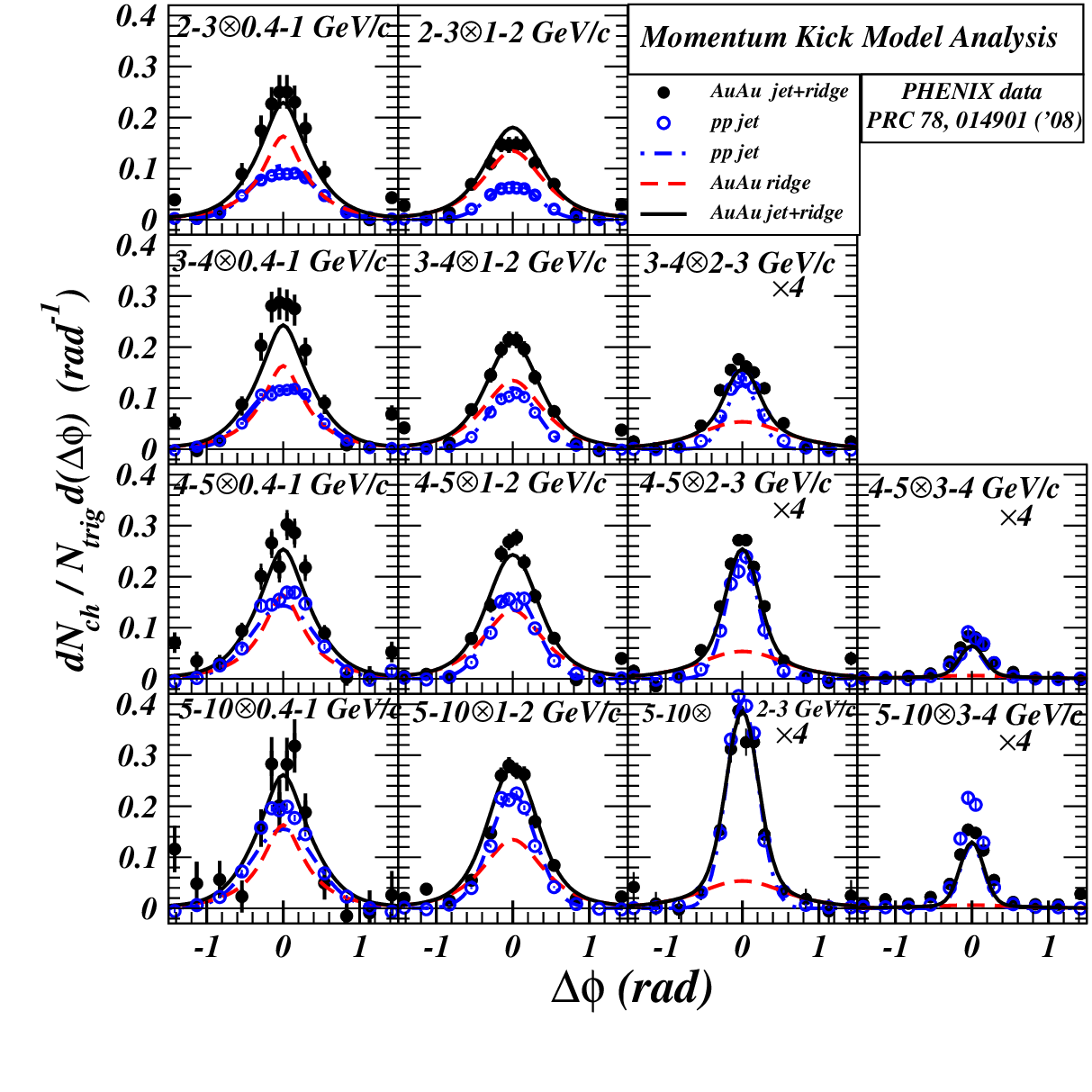}
\vspace*{0.0cm}
\caption{(Color online) PHENIX azimuthal angular distribution of
  associated particles per trigger in different $p_t^{\rm trig}\otimes
  p_t^{\rm assoc}$ combinations.  The solid and open circles are the
  total associated particle yield per trigger, $dN_{\rm ch}/N_{\rm
    trig} d\Delta \phi$, in central Au+Au and $pp$ collisions
  respectively \cite{Ada08}.  The solid, dashed, and dashed-dot curves
  are the theoretical total Au+Au associated particle yields per
  trigger, the Au+Au ridge particle yields per trigger, and the $pp$
  associated particle yields respectively.}
\end{figure}

\vspace*{0.3cm} 
\begin{table}[h]
\caption { Jet component parameters in Eq.\ (\ref{jetfun}) for
associated particles with different $p_t^{\rm trig}$ triggers, in $pp$
collisions at $\sqrt{s_{NN}}$=200 GeV}
\vspace*{0.3cm}
\begin{tabular}{|c|c|c|c|c|c|c|}
\cline{3-7}
 \multicolumn{2}{c|}{}   &  STAR &  \multicolumn{4}{c|} {PHENIX} \\ \hline
Hadron trigger & $p_t^{\rm trig}$ &  4-6GeV & 2-3GeV & 3-4GeV & 4-5GeV  & 5-10GeV  \\ \hline
Properties of  & $N_{\rm jet}$ & 0.75  & 
\multicolumn{4}{c|} {0.15+0.10 $\langle p_t^{trig} \rangle/{\rm GeV}$ } 
\\ \cline{2-7}
particles associated & $T_{\rm jet}$ &{ 0.55{\rm GeV}}  & 
\multicolumn{4}{c|} {0.19 GeV+0.06 $\langle p_t^{trig} \rangle $} 
\\ \cline{2-7}
with a trigger & $\sigma_{\rm \phi 0}$ & \multicolumn {5} {c|} {0.50}    
\\ \cline{2-7}
in $pp$ collisions  & $m_a$ & \multicolumn{5} {c|} {1.1 GeV} \\ \hline
\end{tabular}
\end{table}
\noindent
As indicated in Table II, the parameters of Eqs.\ (\ref{Njetv}) and
(\ref{Tjetv}) are $N_{{\rm jet}0}=0.15$, $d_N=0.1/$GeV, $T_{{\rm
jet}0}=0.19$ GeV, and $ d_T=0.06$.  Thus, particles associated with the
$pp$ collisions changes its properties significantly
as $p_t^{\rm trig}$ changes.

We can compare the parameters obtained here with those from our
previous analysis of the STAR near-side ridge data for central (0-5\%)
Au+Au collisions at $\sqrt{s_{NN}}=200$ GeV, with $4 < p_t^{\rm trig}
<6 $ GeV.  The associated particles in $pp$ collisions in the STAR
measurements can be described by Eq. (\ref{jetfun}) with parameters
\cite{Won08a}
\begin{eqnarray}
\label{jetpar}
N_{\rm jet}=0.67, ~T_{\rm jet}=0.55{\rm ~GeV}, ~\sigma_{\phi 0}=0.50,
 {\rm ~GeV,~and~}~m_a=1.1 {\rm ~GeV},
\end{eqnarray}
as shown in Column 3 of Table II.  They are consistent with
those for the PHENIX associated particle data in $pp$ collisions.

\section{Total Yield of Associated Particles}

The total observed yield of associated particles per trigger in A+A
collisions consists of the sum of the jet and the ridge components,
\begin{eqnarray}
\label{obs}
\left [ 
\frac{1}{N_{\rm trig}}
\frac{dN_{\rm ch}} 
{p_{t} dp_{t} d\Delta \eta  d\Delta \phi } \right ]_{\rm total}^{\rm AA} 
= \left [ f_R   \frac {2}{3} 
\langle N_k \rangle \frac { dF } {p_t dp_t\,
d\Delta \eta\, d\Delta \phi} \right ]_{\rm ridge}^{\rm AA} 
+
\left [ f_J  \frac { dN_{\rm jet}^{pp}} {p_t dp_t\, d\Delta \eta\, d\Delta
\phi} \right ]_{\rm jet}^{\rm AA} \!\!\!\!.
\end{eqnarray}

To obtain the associated ridge yield in the first term on the
right hand side of Eq.\ (\ref{obs}) for A+A collisions, we need
information on the medium parton distribution.  We describe the
normalized initial medium parton momentum distribution, which
implicitly includes all possible physical effects, as represented by
\cite{Won08a}
\begin{eqnarray}
\label{dis2}
\frac{dF}{ p_{ti}dp_{ti}dy_i d\phi_i}&=&
A_{\rm ridge} (1-x)^a 
\frac{ e^ { -\sqrt{m^2+p_{ti}^2}/T }} {\sqrt{m_d^2+p_{ti}^2}},
\end{eqnarray}
where $A_{\rm ridge}$ is a normalization constant defined 
(and determined numerically) by
\begin{eqnarray}
\int dy_i d\phi_i p_{ti}dp_{ti}
{A}_{\rm ridge}  (1-x)^a 
\frac{ \exp \{ -\sqrt{m^2+p_{ti}^2}/T \}} {\sqrt{m_d^2+p_{ti}^2}}
= 1,
\end{eqnarray} 
$x$ is the light-cone variable 
\begin{eqnarray}
\label{xxx}
x=\frac{\sqrt{m^2+p_{ti}^2}}{m_b}e^{|y_i|-y_b},
\end{eqnarray}
$a$ is the fall-off parameter that specifies the rate of decrease of
the distribution as $x$ approaches unity, $y_b$ is the beam parton
rapidity, and $m_b$ is the mass of the beam parton.  A small value of
$a$ indicates a relatively flat rapidity distribution.  In particular,
a boost-invariant rapidity distribution will be characterized by
$a=0$.  A large value of $a\gg 1$ indicates a relatively sharp
fall-off rapidity distribution.  As $x \le 1$, there is a kinematic
boundary that is a function of $y_i$ and $p_{ti}$ at $x=1$,  
\begin{eqnarray}
\label{pty}
\sqrt{m^2+p_{ti}^2}=m_b e^{y_b-|y_i|}.
\end{eqnarray}
We set $m_b$ equal to $m_\pi$ and $y_b$ equal to $y_N$, the rapidity
of the beam nucleons in the CM system.

\vspace*{0.3cm} 
\begin{table}[h]
\caption { Physical parameters in Eqs.\ (\ref{eq2}), (\ref{obser}),
and (\ref{dis2}) in the momentum kick model, and the meaning of each
parameter }
\vspace*{0.3cm}
\begin{tabular}{|c|c|c|}
\hline Category & Physical Parameter & Meaning 
\\ \hline \hline           & $q_L$ 
& magnitude of momentum kick along the jet direction 
\\ 
Properties of & & per jet-(medium parton) collision 
\\ \cline{2-3} 
jet-medium interaction & $f_R\langle N_k\rangle$ & 
centrality-dependent number of kicked partons per trigger 
\\ & & multiplied by the survival factor $f_R$ 
\\ \cline{2-3} 
& $f_J$ & ratio of (jet component yield per trigger in A+A collisions) 
\\ 
& & to (associated jet component in $pp$ collisions) 
\\ \hline \hline

& $a$ & fall-off parameter of medium parton 
\\
Properties of medium parton &  & rapidity distribution in the form $(1-x)^a$
\\ \cline{2-3} 
momentum distribution in & $T$ & ``temperature'' of the medium 
parton $p_t$ distribution
\\ \cline{2-3} 
central A+A Collisions & $m_d$ & mass parameter to
modify the $p_t$ distribution for low $p_t$ 
\\ \hline
\end{tabular}
\end{table}
\vspace*{0.3cm}

From the above discussions, we note that the momentum kick model
physical parameters can be divided into two categories as given in
Table III where the meaning of each parameter is listed.  There are
parameters $q_L$, $f_R\langle N_k\rangle$, and $f_J$ which pertain to
the jet-medium interaction.  They provide information on the momentum
kick per collision $q_L$ along the jet direction, the number of
jet-(medium parton) collisions $\langle N_k \rangle$ multiplied by
$f_R$, and the ratio $f_J$ of the jet component in A+A collisions per
trigger relative to the jet component in $pp$ collisions.  Finally,
there are parameters $a$, $T$, and $m_d$ which pertains to the
properties of the medium at the moments of jet-(medium parton)
collisions.  They provide information on the shape of the early medium
parton momentum distribution.  The evaluation of these quantities from
fundamental theories is beyond the scope of the present theoretical
development.  

In calculating theoretical differential distribution $dN_{\rm
  ch}/N_{\rm trig} d\Delta \eta$ as a function of $\Delta \eta$, we
  impose the experimental constraints of $\eta_{\rm min}^{\rm trig}
  \le \eta^{\rm trig} \le \eta_{\rm max}^{\rm trig}$ and $\eta_{\rm
  min}^{\rm assoc} \le \eta^{\rm assoc} \le \eta_{\rm max}^{\rm
  assoc}$ which generate various pseudorapidity differences $\Delta
  \eta=\eta^{\rm assoc}-\eta^{\rm trig}$.  We add up all yields
  $dN_{\rm ch}/N_{\rm trig} d\Delta \eta$ of the same $\Delta
  \eta=\eta^{\rm assoc}-\eta^{\rm trig}$, to get the uncorrected yield
  as a function of $\Delta \eta$.  We assume that the acceptance is
  uniform in regions of both $\eta^{\rm assoc}$ and $\eta^{\rm trig}$.
  Theoretical acceptance-corrected yield is then equal to the product
  of the uncorrected yield and the acceptance correction factor
  $f_{\rm acc}(\Delta \eta)$.  We can alternatively carry out the
  acceptance correction as the uncorrected yield divided by the factor
  $[1/f_{\rm acc}(\Delta \eta)]$ arising from a uniformly generated
  distribution in $\eta^{\rm trig}$ and $\eta^{\rm assoc}$.

The acceptance correction factor $f_{\rm acc}(\Delta \eta)$ can be
obtained from geometrical considerations by plotting the acceptance
region in the plane of $\eta^{\rm assoc}$ and $\eta^{\rm trig}$ and
changing the axes to $\eta^{\rm assoc}-\eta^{\rm trig}$ and $\eta^{\rm
  assoc}+\eta^{\rm trig}$.  From the geometrical areas after the
change of axes, the $\Delta \eta$ acceptance correction factor is
given by
\begin{eqnarray}
\label{facc}
f_{\rm acc}(\Delta \eta) =\begin{cases}
\frac{(\eta^{\rm trig}_{\rm max}-\eta^{\rm trig}_{\rm min})}
{\Delta \eta - (\eta^{\rm assoc}_{\rm min}-\eta^{\rm trig}_{\rm max}) } 
  & {\rm ~for~} \eta^{\rm assoc}_{\rm min}-\eta^{\rm trig}_{\rm max} 
  < \Delta \eta \le  \eta^{\rm assoc}_{\rm min}-\eta^{\rm trig}_{\rm min}\\
~~~~~~~1 & {\rm ~for~} \eta^{\rm assoc}_{\rm min}-\eta^{\rm trig}_{\rm min} 
  \le \Delta \eta \le  \eta^{\rm assoc}_{\rm max}-\eta^{\rm trig}_{\rm max}\\
\frac{(\eta^{\rm trig}_{\rm max}-\eta^{\rm trig}_{\rm min})}
{(\eta^{\rm assoc}_{\rm max}-\eta^{\rm trig}_{\rm min})-\Delta \eta } 
  & {\rm ~for~} \eta^{\rm assoc}_{\rm max}-\eta^{\rm trig}_{\rm max} 
  \le \Delta \eta <  \eta^{\rm assoc}_{\rm max}-\eta^{\rm trig}_{\rm min}.\\
    \end{cases}
\end{eqnarray}
A computer program to carry out the momentum kick model
analysis outlined above can be obtained from the author upon request.

\section{Analysis of PHENIX Ridge Data}

In this Section, we investigate the PHENIX near-side ridge data for
$pp$ collisions and for the most central (0-20\%) Au+Au collisions at
$\sqrt{s_{NN}}=200$ GeV.  The region of
acceptance includes $|\eta^{\rm trig},\eta^{\rm assoc} | < 0.35$ and
many $p_t^{\rm trig} \otimes p_t^{\rm assoc}$ combinations
\cite{Ada08}.

We need to find out how the jet component in Au+Au collisions is
related to the jet component in $pp$ collisions.  Previously, the jet
component per trigger in Au+Au collisions can be considered as an
attenuated jet component in $pp$ collisions with a survival factor
$f_J \sim 0.632$ \cite{Won08a}.  For the PHENIX experimental data, we
find that $f_J$ increases to unity as $p_t^{\rm assoc}$ increases to
3-4 GeV.  We can understand this behavior because the jets with
$p_t^{\rm assoc}>3-4$ GeV are likely to come from the fragmentation
process outside the medium and the associated particles are likely to
be unattenuated.  As $f_J=0.632$ for $p_t^{\rm assoc}$$<$2 GeV and 
$f_J=1.0$ for $p_t^{\rm assoc}$$>$3 GeV respectively, we can interpolate
$f_J=0.82$ in the intermediate region and use an empirical $f_J$
factor that depends on $p_t^{\rm assoc}$,
\begin{eqnarray}
\label{fj}
f_J(p_t^{\rm assoc})=\begin{cases}
0.632 & {\rm for~}p_t^{\rm assoc}<2 {\rm ~GeV},   \\
0.82  & {\rm for~}2<p_t^{\rm assoc}<{\rm3~GeV},  \\
1.0   & {\rm for~3~GeV} <p_t^{\rm assoc} . \\
\end{cases}
\end{eqnarray}

With the knowledge of jet component in Au+Au collisions, we can
determine the properties of the medium and the characteristics of the
jet-medium interaction.  For the medium momentum distribution given by
Eq.\ (\ref{dis2}), we use the same shape as that obtained previously
in the analysis of the STAR ridge data \cite{Won08a} with parameters
\begin{eqnarray}
a = 0.5, ~T = 0.50 {\rm ~~GeV,~ and~~} m_d = 1 {\rm ~GeV}.
\end{eqnarray}
The remaining free parameters are then the magnitude of the momentum
kick $q_L$ along the jet direction and the number of attenuated medium
kicked partons $f_R\langle N_k\rangle$.  The PHENIX ridge data are
found to be well described (Fig.\ 1) by
\begin{eqnarray}
q_L = 0.8 {\rm ~~GeV,~ and~~} f_R\langle N_k\rangle= 3.0.
\end{eqnarray}
We summarize the values of the parameters for the analysis of the
PHENIX ridge data in Table IV.  As a comparison, we also list the
values of the parameters used previously in the analysis of the STAR
and PHOBOS ridge data.
\begin{table}[h]
\hspace*{2.0cm} 
\caption{Jet-medium interaction and medium parton distribution
  parameters in Eqs.\ (\ref{eq2}) and (\ref{dis2}) in the momentum
  kick model for particles associated with a hadron trigger with
  different $p_t^{\rm trig}$ in central Au+Au Collisions at
  $\sqrt{s_{NN}}$=200 GeV}
\vspace*{0.3cm}
\begin{tabular}{|c|c|c|c|c|c|c|c|}
\cline{3-8} \multicolumn{2}{c|}{} & STAR & PHOBOS & \multicolumn{4}{c|} {PHENIX} \\ \hline 
\multicolumn{2} {|c|} {\hspace*{-0.8cm}Centrality} & 0-5\% & 0-10\% &
            \multicolumn{4} {c|} {0-20\%} \\ \hline 
Hadron trigger & $p_t^{\rm trig}$ &  4-6GeV & $>$2.5GeV & 2-3GeV 
                 & 3-4GeV & 4-5GeV & 5-10GeV \\ \hline 
Momentum kick & $q_L$ & \multicolumn{2} {c|}{1.0 GeV} 
           & \multicolumn{4} {c|} {\hspace*{-0.0cm}0.80 GeV} \\ \hline 
Number kicked partons & $f_R\langle N_k\rangle$ 
           & \multicolumn{2} {c|} {3.8} 
           & \multicolumn{4} {c|} {\hspace*{-0.0cm}3.0} \\ \hline 
Jet component &  $f_J$ & \multicolumn{2} {c|} {0.632} &
\multicolumn{4} {c|} {0.632 for $p_t^{\rm assoc}<$2 GeV} \\
survival factor &  &  \multicolumn{2} {c|} {} &
\multicolumn{4} {c|} {~~~~ 0.82~~ for 2 $<p_t^{\rm assoc}<$3 GeV}\\
                &  &   \multicolumn{2} {c|} {} &
\multicolumn{4} {c|} {\!1.00~ for 3 GeV$< p_t^{\rm assoc}$}  \\ \hline\hline 
Medium  parton & $a$ & \multicolumn{6} {c|}{0.5} \\ \cline{2-8}
distribution parameters in & $T$ & \multicolumn{6} {c|}{0.5 GeV} \\ \cline{2-8} 
central Au+Au Collisions & $m_d$ & \multicolumn{6} {c|} {1.0 GeV} \\ \hline
\end{tabular}
\end{table}

In Fig.\ 1, we show the PHENIX near-side ridge data \cite{Ada08} for
collisions at $\sqrt{s_{NN}}=200$ GeV and the momentum kick model
theoretical results.  The solid data point are the total associated
particle yield per trigger, $dN_{\rm ch}/N_{\rm trig} d\Delta \phi$,
in central Au+Au collisions, and the open circles are the associated
particle yields per trigger, $dN_{\rm ch}/N_{\rm trig} d\Delta \phi$,
in $pp$ collisions \cite{Ada08}.  Theoretical results are given as
various curves.  The solid, dashed, and dashed-dot curves are the
theoretical total Au+Au associated particle yields per trigger, the
Au+Au ridge particle yields per trigger, and the $pp$ associated
particle yields, respectively.  The different subfigures give the
yields of associated particles with different $p_t^{\rm assoc}$,
spanning $p_t^{\rm assoc}$ from 0.4 GeV up to $p_t^{\rm trig}$.

Comparison of the PHENIX near-side data with the results of the
momentum kick model in Figs.\ 1 indicates that the PHENIX ridge data
on the near-side for central Au+Au collisions at $\sqrt{s_{NN}}=200$
GeV \cite{Ada08} can be well described by the momentum kick model.

We can compare the extracted values of physical parameters of the
jet-medium interaction and medium parton characteristics with those
extracted previously from the STAR data.  The centrality region
covered by the present PHENIX measurement \cite{Ada08} extends from 0
to 20\%, whereas the STAR data \cite{Ada05} extends from 0 to 5\%.
The method of subtracting the $v_2$ background are also different
\cite{Jia08a}.  The STAR detector covers $|\eta|<1$ and
0$<$$\phi$$<$$2\pi$; the PHENIX detector covers $|\eta|< 0.35$ and
only about half of the full range of azimuthal angles.  The medium
parton parameters $a$, $T$, and $m_d$ are the same, whereas $(q_L,
f_R\langle N_k\rangle)$ is (1 GeV, 3.8) for the STAR data and (0.8
GeV, 3.0) for the PHENIX data.  The difference in $q_L$ and $f_R\langle
N_k\rangle$ may arise from difference in centrality selections and the
methods of processing the data.

From the present study of the PHENIX ridge data in the region of
$|\eta| < 0.35$, we can briefly compare the associated particle yield
per trigger of the ridge component and the jet component in central
Au+Au collisions, as a function of $p_t^{\rm assoc}$.  We find from
Figs.\ 1 that the ridge associated particle yield per trigger is
comparable to the jet associated particle yield for $p_t^{\rm assoc}
\lesssim 2$ GeV.  Thus, ridge particles show up as an excess to the
jet component in the region of small $\Delta \eta$ and $ \Delta \phi
\sim 0$, for $p_t^{\rm assoc} \lesssim 2$ GeV.  However, for $p_t^{\rm
assoc}$$>$2-3 GeV, the jet component dominates over the ridge
component.  This variation of the relative strengths of the jet and
ridge components is reproduced well by the momentum kick model.  The
physical reason for the large contribution of the ridge component
around $p_t$$\sim$1 GeV arises from fact that the ridge momentum
distribution is in fact the initial transverse momentum distribution
shifted by a momentum of about 1 GeV.

The range of $\Delta \eta$ examined by the PHENIX Collaboration is
relatively small.  A much larger range of $\Delta \eta$ has been
investigated by the STAR and PHOBOS Collaborations.  As a function of
$\Delta \eta$, the jet component decreases rapidly away from the peak
at $(\Delta \phi, \Delta \eta)$$\sim$ 0, whereas the ridge component
extends to regions of large $|\Delta \eta|$ and it dominates over the
jet component at $|\Delta \eta|$, as observed by the STAR \cite{Wan07}
and PHOBOS Collaborations \cite{Wen08}.  This feature in the variation
in $\Delta \eta$ is also reproduced by the momentum kick model
\cite{Won08a}.

As the jet component in Au+Au collisions per trigger is related to the
associated particles in $pp$ collisions, and the characteristics of
the associated particles in $pp$ collision change significantly as
$p_t^{\rm trig}$ changes, so the jet component per trigger in the
Au+Au collision also changes its properties significantly as $p_t^{\rm
  trig}$ changes.  The temperature $T_{\rm jet}$ and the number of
these associated particles $N_{\rm jet}$ increases linearly with
$p_t^{\rm trig}$.

In contrast to the large variation of the properties of the jet
component as a function of $p_t^{\rm trig}$, physical parameters
associated with the medium partons appears to be relatively robust,
independent of $p_t^{\rm trig}$.  The same set of medium property
parameters of $a$, $T$, and $m_d$ apply to the medium parton momentum
distribution, for all $p_t^{\rm trig}$ and $p_t^{\rm assoc}$
combinations.  They coincide also with those from STAR and PHOBOS
measurements \cite{Won08a}. The robust nature of these physical
quantities enhances their quality as basic properties of the produced
medium.  The fall-off parameter $a=0.5$ for the distribution $(1-x)^a$
of Eq.\ (\ref{dis2}) reveals that the early medium parton rapidity
distribution is relatively flat but not boost-invariant, which
would correspond to $a=0$.  The $(1-x)^a$ distribution with the
kinematic limit of $x=1$ indicates that the distribution is in the
shape of a rapidity plateau, as shown in Fig.\ 6(b) of \cite{Won08a}.
The temperature parameter $T=0.5$ GeV shows that it is a thermal-like
distribution with a temperature between those of a high-$p_t$ jet and
the bulk matter.  The quantity $m_d=1$ GeV indicates a small
modification of the thermal distribution at lower $p_t$.

Similarly, the set of physical parameters that describe the jet-medium
interaction, $q_L$ and $f_R \langle N_K\rangle$, appear also to be
robust as the same set can describe the ridge component for all
different $p_t^{\rm trig}$ and $p_t^{\rm assoc}$ combinations.  The
extracted magnitude of the momentum kick is $q_L=0.8$ GeV per
jet-(medium parton) collision, and the number of jet-medium parton
collision for the most central collisions multiplied by the
survival factor is 3.

There is however a difference of about 20\% in the values of $q_L$ and
$f_R \langle N_K\rangle$ extracted from the PHENIX near-side ridge
data, compared to those extracted from the STAR near-side ridge data.
This difference may reflect the difference in centrality selection and
the degree of uncertainty in processing the experimental data.

\section{The nature of the scattering between the jet parton and the
  medium parton}

We have extracted the relevant physical quantities from the ridge
data.  We come to the second stage of our analysis to find out the
nature of the collision between the jet parton and the medium parton.
We also wish to correlate the extracted physical quantities to those
in relevant physical phenomena to see whether they are consistent.  We
shall discuss the extracted magnitude of the momentum kick $q_L$ in this
Section, the extracted shape of the rapidity distribution in Section
VII, and the extracted number of kicked partons in Section VIII.

The extracted magnitude of the momentum kick $q_L$ is the longitudinal
momentum imparted by the jet parton onto the medium parton per
collision, along the jet direction. This quantity $q_L$ is also the
longitudinal momentum loss of the incident jet parton in the
parton-parton collision.  We find $q_L=$0.8 GeV for the present set of
PHENIX data with 20\% centrality and $q_L=$1.0 GeV previously for the
STAR ridge data with 0-5\% centrality.  The average $q_L$ value from the
two measurements is $q_L=0.9$ GeV.

To study the scattering between the jet parton and the medium parton,
we would like to relate the longitudinal momentum loss $q_L$ of the jet
to its momentum transfer squared $t$.  The latter quantity is related
to the scattering correlation length $a$ in an elastic parton-parton
collision, for which many pieces of information have been obtained
previously
\cite{Lan88,Nac91,Dos87,Kra90,Dos92,Dos01,Ber99,DiG92,Meg99,For97,Don02,Sch81,Cho68}.

It is convenient to work in the medium rest frame in which the average
velocity of the medium partons is zero.  We consider the collision of
an energetic jet parton $a$ with an medium parton $b$ at rest, which
represents an average parton of the medium.  For simplicity, we
specialize to the case in which all partons have the same rest mass
$m$.  The elastic scattering of the jet parton $a$ with the medium
parton $b$ leads to partons $c$ and $d$ as
\begin{eqnarray}
a + b \to c + d. 
\end{eqnarray}
The square of the center-of-mass energy of the colliding 
partons is
\begin{eqnarray}
s = (a+b)^2= 2 m^2 + 2 m \sqrt{{\bf a}^2 + m^2} ,
\end{eqnarray}
where we have used the same label for a parton and its 3- and
4-momentum.  In the elastic scattering, the momentum transfer squared
$t=(a-c)^2$ is
\begin{eqnarray}
t=-\frac{1}{2}(s-4m^2)(1-\cos \theta^*)
\end{eqnarray}
where $\theta^*$ is the scattering angle between ${\bf c}^*$ and ${\bf
a}^*$, and the superscript $*$ denotes quantities in the $(a+b)$
center-of-mass system.  There is thus a relation between the
scattering angle $\theta^*$ and the momentum transfer $t$.  The
maximum and minimum values of $t$ are
\begin{eqnarray}
t_{\rm max}=0,          & &~~ {\rm for~~} \theta^*=0,
\nonumber\\
t_{\rm min}=-(s-4m^2),  & &~~ {\rm for~~} \theta^*=\pi.
\end{eqnarray}

After the elastic parton-parton scattering, parton $a$ becomes parton
$c$ with the same energy and the magnitude of the 3-momentum,
\begin{eqnarray}
|{\bf c}^*| = \frac{1}{2} \sqrt{s-4m^2}.
\end{eqnarray}
The longitudinal component of $c$  in the center-of-mass system
is
\begin{eqnarray}
c_z^* (\theta^*) = \frac{1}{2} \sqrt{s-4m^2} \cos\theta^*.
\end{eqnarray}
Transforming back to the medium rest frame, one obtains the
longitudinal momentum $c_z$ of the final parton in the medium rest
frame to be
\begin{eqnarray}
c_z(\theta^*) =\gamma [ c_z^* (\theta^*) +\beta c_0^*],
\end{eqnarray}
where the Lorentz transformation factors $\gamma$ and $\beta$ are
$\gamma=\sqrt{s}/2m$ and $\beta=\sqrt{1-\gamma^{-2}}$.  The final
momentum of $c_z$ for the case of $\theta^*=0$ gives the initial
longitudinal momentum of the incident parton $a_z =c_z(\theta^* =0)$.
Therefore, the momentum loss of the incident jet parton in the medium
rest frame is
\begin{eqnarray}
\label{qqq}
q_L=c_z(\theta^*=0) - c_z(\theta^*) = -\frac{t}{2m}
\frac {\sqrt{s}}{\sqrt{s-4m^2}}.
\end{eqnarray}
In the elastic parton-parton scattering, the longitudinal momentum
loss $q_L$ of the incident jet parton is equal to the longitudinal
momentum gain or momentum kick $q_L$ suffered by the medium parton along
the jet direction.  Thus, for a given incident parton with a definite
parton-parton center-of-mass $\sqrt{s}$, the knowledge of the
magnitude of the momentum kick $q_L$ will provide information on the
momentum transfer squared $|t|$.

In detecting a trigger of energy $4$-$6$ GeV, the incident jet parton
has an energy of order 10 GeV, as the partons loses about a few GeV in
kicking a few medium partons.  For this incident parton momentum of
order $p_t^{\rm jet}\sim$10 GeV in our present experimental setup,
$\sqrt{s} \gg m$ and $q\sim |t|/2m$, which is independent of the
parton energy.  For $q_L$=0.9 GeV extracted from the momentum kick
model, we therefore obtain the squared momentum transfer $t$ to have
the magnitude
\begin{eqnarray}
\label{tval}
|t| = 2mq_L\frac{\sqrt{s-4m^2}}{\sqrt{s}} = 0.255 {\rm ~~GeV}^2. 
\end{eqnarray}
As the longitudinal momentum gained by the medium parton in the
momentum kick model is an average quantity, the corresponding $t$ in
Eq.\ (\ref{tval}) should be considered as an average value $\langle
|t|\rangle$.

The extracted value of the (average) momentum transfer squared $|t|$
in the parton-parton collision is small, substantially less than 1
GeV$^2$.  This suggests that the collision process is
non-perturbative.  The parton-parton scattering should be more
appropriately described by the exchange of a non-perturbative pomeron
\cite{Lan88,Nac91,Dos87,Kra90,Dos92,Dos01,Ber99,DiG92,Meg99,Don02,For97}.

We would like to relate the (average) momentum transfer squared $t$ to
a correlation length $a$ by considering a parton-parton collision
profile function of the form
\begin{eqnarray}
\Gamma({\bf b})=\frac{\Gamma_0}{2\pi a^2} \exp\{ - \frac{{\bf b}^2}{2 a^2} \}
\end{eqnarray}
where $\Gamma_0$ is the scattering strength parameter.  The scattering
amplitude is
\begin{eqnarray}
f({\bf q}_t) &=&\frac{ik}{2\pi} \int d{\bf b} 
%\exp \{i {\bf q}_t \cdot {\bf b}\}
e^{i {\bf q}_t \cdot {\bf b}}
\Gamma({\bf b})
\nonumber\\
&=&\frac{ik}{2\pi} 
\frac{\Gamma_0 a^2}{2\pi } \exp \{ - \frac{a^2 {\bf q}_t^2}{2 } \}.
\end{eqnarray}
As ${\bf q}_t^2=-t-t^2/(s-4m^2) \sim -t$, the elastic parton-parton
scattering differential cross section is 
\begin{eqnarray}
\label{exp}
\frac{d\sigma}{dt}
\sim \frac{\Gamma_0 a^4 }{8 \pi^2} e^{a^2 t}. 
\end{eqnarray}
The average value of $|t|$ is therefore
\begin{eqnarray}
\label{ttt}
\langle~|t|~\rangle = \int_{t_{\rm min}}^{t_{\rm max}} |t| \frac{d\sigma}{dt}
\biggl / \int_{t_{\rm min}}^{t_{\rm max}}  \frac{d\sigma}{dt}
\sim \frac{1}{a^2},
\end{eqnarray}
which allows us to infer the magnitude of the correlation length $a$
from the average value of momentum transfer squared $\langle |t|
\rangle $.  From Eq.\ (\ref{ttt}) and (\ref{tval}), the magnitude of
the longitudinal momentum kick $q_L$ extracted from the ridge data
corresponds to a parton-parton scattering correlation length $a$ of
\begin{eqnarray}
\label{cor}
a \sim  1/ \sqrt{ \langle ~|t| ~\rangle} =
0.39 {\rm ~~fm}.
\end{eqnarray}

Is this correlation length $a$ compatible with measurements of the
same quantity in other descriptions of the parton-parton elastic
collision process?  One can consider a model of hadron-hadron
collisions in which the partons of one hadron collide with partons of
the other hadron, as in the Chou-Yang model \cite{Cho68}.  In the
original Chou-Yang droplet model \cite{Cho68}, the partons are assumed to be
point-like without any structure of a correlation length.  The
Chou-Yang model of the point-like parton-parton scattering can be
generalized to the case of partons with a finite correlation length,
with the parton-parton scattering differential cross section assuming
the form of Eq.\ (\ref{exp}) \cite{Sch81,Bia76,Lev74}.  The elastic
hadron-hadron elastic differential cross section in the modified
Chou-Yang model then takes the form \cite{Sch81}
\begin{eqnarray}
\label{Chouyang}
\frac{d\sigma_{\rm hadron-hadron}}{dt}
=A F_p^2 (t) F_t^2 (t) |A_{qq}(t)|^2,
\end{eqnarray}
where $A$ is a normalization factor, $F_p(t) $ and $F_t(t)$ are the
projectile and target hadron form factors respectively, and
$|A_{qq}(t)|^2$ is the quark-quark scattering matrix element taken to
have the same functional form as Eq.\ (\ref{exp}),
\cite{Sch81,Bia76,Lev74}
\begin{eqnarray}
|A_{qq}(t)|^2 = e^{a^2 t}.
\end{eqnarray}
Experimental $pp$,
$\pi^+p$, and $\pi^- p$ elastic differential cross sections at 200 GeV
can be well described by 
\begin{eqnarray}
\label{aChou}
a =\begin{cases}
   0.33 & {\rm ~fm~for~} pp    {\rm ~collisions}; \cr 
   0.25 & {\rm ~fm~for~} \pi p {\rm ~collisions}.  \cr
   \end{cases}
\end{eqnarray}
(see Fig.\ 14 and Table X of \cite{Sch81}, where the correlation
length $a$ is represented in terms of the ``quark radius $r_q$'' with
$r_q=\sqrt{2} a$.)  

It is of interest to inquire further whether the correlation length
(\ref{cor}) extracted from the ridge data is compatible with the
correlation length in the non-perturbative description of the pomeron,
for which much progress has been made in recent years. The slow rise
of the total hadron-hadron cross sections with increasing energy as
$(\sqrt{s})^{0.0808}$ in high energies hadron-hadron collisions
suggests that the scattering process is dominated by the exchange of a
pomeron, whose quantum numbers are those of the vacuum
\cite{Don02,For97,Won94}.  The approximate validity of the additive
quark model, where the cross section is proportional to the valence
quark number, suggests that the exchange of the pomeron takes place as
the exchange between single quark partons.

In QCD, it is natural to assume that the exchange of the pomeron
between constituent quark partons is just the exchange of
a cluster of two or more gluons, in order to get the correct quantum
number of the vacuum \cite{Low75,Dol92,Zha03}.  In perturbative QCD,
the perturbative exchange of two gluons between quark partons leads to
a singularity of the elastic hadron-hadron scattering amplitude at
$t=0$, and it does not reproduce the experimental $t$-dependence.  The
experimental differential cross section, $d\sigma_{\rm
hadron-hadron}/dt$ corresponds more properly to the hadron form
factors, as in Eq.\ (\ref{Chouyang}), with the cluster of exchanged
gluons coupled to a single quark.  It is more appropriate to describe
the exchange of the pomeron to be a non-perturbative process and take
into account non-perturbative properties of the QCD vacuum.

The non-perturbative QCD vacuum can be described as consisting of a
gluon condensate of a color field strength characterized by
\cite{Shi79}
\begin{eqnarray}
\langle g^2 F_{\mu \nu}^C(0) 
F^{C,\mu \nu}(0)\rangle_A = M_c^4, 
\end{eqnarray}
where the expectation value is taken with respect to the
non-perturbative vacuum and $M_c=(0.9 \pm 0.1 {\rm~GeV})$
\cite{Nar95}.  For the description of the pomeron in the scattering
process, Landshoff and Nachtmann \cite{Lan88,Nac91} generalized the
concept of the gluon condensate to the case of a gluon condensate with
a finite correlation length $a$ associated with each colliding
parton,
\begin{eqnarray}
\langle g^2 F_{\mu \nu}^C(x) 
F^{C,\mu \nu}(y)\rangle_A = M_c^4 f((x-y)^2/a^2). 
\end{eqnarray}
where $x-y$ is space-like with $(x-y)^2 <0$.  From hadron spectroscopy 
and the total cross section at high energies, the correlation length was estimated to be \cite{Kra90,Nac91}
\begin{eqnarray}
\label{Kra}
a\approx 0.4 {\rm ~fm}.
\end{eqnarray}
A parton-parton scattering is then described as taking place by the
scattering of the parton on the condensate that is associated with the
collided partons, as in potential scattering.  

The concept of correlators with a correlation length was further
developed in QCD in the Model of Stochastic Vacuum (MSV) \cite{Dos87}.
In this model, the gluon condensate correlator is assumed to be
described by invariant functions $D$ and $D_1$ which are normalized to
$D(0)=D_1(0)=1$. They are given explicitly in \cite{Dos92,Ber99} and
they fall off rapidly on a scale of the correlation length $a$.  The
correlators are separated into a part of strength $\kappa$ that is
non-Abelian and a part of strength $(1-\kappa)$ that is Abelian
\cite{Dos92,Dos01},
\begin{eqnarray}
\langle g^2 F_{\mu \nu}^C(x)  F_{\rho \sigma}^D(y) \rangle
&=&\delta^{CD} \frac{ \pi^2 G_2}{6}
 \bigg \{ \kappa (\delta_{\mu \rho}  \delta_{\nu \sigma} 
                - \delta_{\mu \sigma}\delta_{\nu \rho}) D(\frac{(x-y)^2}{a^2})
 \nonumber \\
&+& (1- \kappa) \frac{1}{2} [  
     \frac{\partial}{\partial z_\mu}(z_{ \rho}  \delta_{\nu \sigma} 
                - z_{\sigma}\delta_{\nu \rho}) 
    +\frac{\partial}{\partial z_\nu}(z_{ \sigma}\delta_{\nu \rho} 
                - z_{\rho}  \delta_{\nu \sigma})] D_1(\frac{(x-y)^2}{a^2}),
\end{eqnarray}
where $C$ and $D$ are color indices, and $\pi^2
G_2/6=M_c^4$. Differential hadron-hadron cross sections for
high-energy elastic scattering have been analyzed and the experimental
data can be described well by using a condensate correlator with
parameters \cite{Ber99}
\begin{eqnarray}
\label{par}
a& =&0.32 {\rm ~fm},
\nonumber\\
\kappa& = &0.74,
\nonumber\\
{\rm ~and~}
G_2 &=& (0.529 {\rm ~GeV})^4.
\end{eqnarray}
The properties of the gluon condensate can be further examined by
lattice gauge calculations.  The correlation length parameter $a$, the
non-Abelian parameter $\kappa$, and the gluon condensate strength
$G_2$ in Eq. (\ref{par}) obtained in quenched lattice calculations
\cite{DiG92,Meg99} are compatible with those obtained in hadron-hadron
differential cross sections \cite{Ber99}.

We can summarize the values of the correlation length $a$ in a
parton-parton collision from different investigations in Table V.

\vspace*{0.3cm} 
\begin{table}[h]
  \caption { Comparison of the correlation length $a$ obtained from 
various considerations}
\vspace*{0.3cm}
\begin{tabular}{|c|c|c|}
  \hline Source & Correlation Length $a$ (fm) & References 
  \\ \hline \hline   
  Momentum Kick Model  & 0.39 &  Present investigation
  \\ \hline
  Small $|t|$ hadron-hadron $d\sigma/dt$  & 0.33 ($pp$) & \cite{Sch81}
  \\ \cline{2-3} 
  in modified Chou-Yang Model & 0.25 ($\pi p)$  & \cite{Sch81}
  \\ \hline
  Hadron spectroscopy \& $\sigma_{\rm tot}$ & $\approx$ 0.40         &  \cite{Kra90,Nac91}
  \\ 
  in Stochastic Vacuum Model   &           & 
  \\ \hline
  Small $|t|$ hadron-hadron $d\sigma/dt$  & 0.32         &  \cite{Ber99}
  \\ 
  in Stochastic Vacuum Model   &           & 
  \\ \hline
  Gluon correlators               & 0.22-0.48 &  \cite{Meg99}
  \\ 
  in lattice gauge calculations    &           & 
  \\ \hline
\end{tabular}
\end{table}

It is gratifying that the correlation length estimated from the ridge
data, $a=0.39$ in Eq.\ (\ref{cor}), is compatible with the value of
$a=0.25$-0.33 fm in Eq.\ (\ref{aChou}), obtained in a hadron-hadron
elastic scattering in the modified Chou-Yang model \cite{Sch81}, the
value of $a\approx 0.4$ fm in Eq.\ (\ref{Kra}), obtained from hadron
spectroscopy and total cross sections in the Model of Stochastic Vacuum
\cite{Kra90}, the value of $a=0.32$ fm in Eq.\ (\ref{par}), obtained in
the non-perturbative pomeron description in the Model of Stochastic
Vacuum \cite{Ber99}, and the value of $a=0.22$-0.48 fm obtained in
lattice gauge calculations \cite{Meg99}.  The approximate
agreement of the correlation length extracted in the momentum kick
model (\ref{cor}) with many previous results supports the approximate
validity of the magnitude of the momentum kick $q_L$ in the present
analysis.

Because of the small value of $|t|< 1$ GeV and the compatibility of of
the correlation length $a$ with previous description of the
non-perturbative pomeron, we conclude that the parton-parton
scattering between the jet parton and the medium parton arises from
the exchange of a non-perturbative pomeron, in the momentum range of
$p_t^{\rm trig}<10$ GeV considered in the near-side ridge
measurements.  Our ability to ascertain the nature of the
parton-parton scattering will help us select the proper description to
formulate the process of energy loss for these jet partons.

\section{The Occurrence of the Rapidity Plateau}

The initial momentum distribution of the medium partons in
Eq. (\ref{dis2}), $(1-x)^a\exp\{ - \sqrt{m^2+p_t^2}/T \}/\sqrt{m_d^2+
p_t^2}$, gives an early parton momentum distribution which has three
prominent features.  First, it has a thermal-like transverse
distribution whose characteristic slope parameter $T$ is between those
of the jet and the bulk inclusive matter.  Second, the rapidity
distribution is relatively flat around $y\sim 0$.  Third, the rapidity
distribution is quite extended, reaching out to large rapidities.

Questions may be raised with regard to the occurrence of the rapidity
plateau in the early parton momentum distribution.  One immediate
question is whether such a plateau structure also occurs in the
distribution of produced particles in related phenomena.
Theoretically, the rapidity plateau occurs in elementary processes
involving the fragmentation of flux tubes \cite{
  Cas74,Bjo83,Won91,Won94} and in many particle production models such
as models based on preconfinement \cite{Wol80}, parton-hadron duality
\cite{Van88} cluster fragmentation \cite{Odo80}, string-fragmentation
\cite{And83}, dual-partons \cite{Cap78}, the Venus model \cite{Wer88},
the RQMD model \cite{Sor89}, multiple collision model \cite{Won89},
parton cascade model \cite{Wan94,Gei92}, color-glass condensate model
\cite{McL94}, the AMPT model \cite{Li96}, the Lexus model
\cite{Jeo97}, and many other models.  To investigate the origin of the
rapidity plateau in a quantum mechanical framework, we can go a step
further to use the physical argument of transverse confinement to
establish a connection between QCD and QED2 (Quantum Electrodynamics
in 2-dimensions) \cite{Won08a,Won09}.  One finds that a rapidity
plateau of produced particles is a natural occurrence when color
charges pull away from each other at high energies
\cite{Cas74,Bjo83,Won91,Won94} as in QED2
\cite{Sch62,Low71,Col75,Col76}.  Experimental evidence for a plateau
in rapidity distributions along the sphericity axis or the thrust axis
has been observed earlier in $\pi^\pm$ production in high-energy
$e^+$-$e^-$ annihilation \cite{Aih88,Hof88,Pet88,Abe99,Abr99} and in
$pp$ collisions at RHIC energies \cite{Yan08}.

\begin{figure} [h]
\includegraphics[angle=0,scale=0.60]{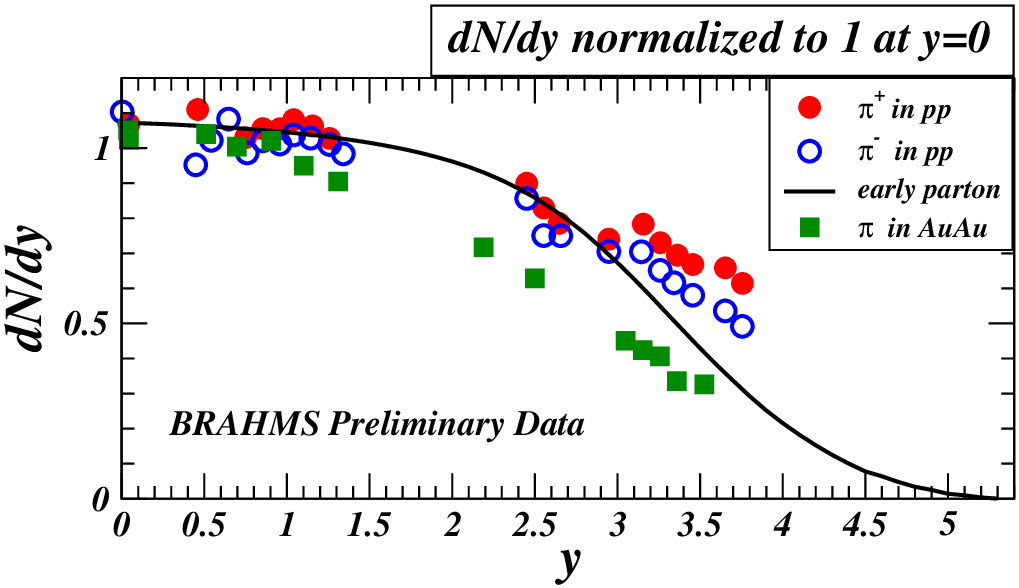}
\vspace*{0.0cm}
\caption{(Color online) {The data points are the pion rapidity
distributions for $pp$ and Au+Au collisions from the BRAHMS
Collaboration.  The solid curve is the early parton momentum
rapidity distribution extracted from the ridge data associated with
the near-side jet.}}
\end{figure}

To gain a new insight into the rapidity plateau structure of the early
partons, it is of interest to compare the shape of the rapidity
distribution of the early partons extracted in the momentum kick model
with those from $pp$ and central Au+Au collisions at the same energy,
$\sqrt{s_{NN}}=200$ GeV.  We plot in Fig.\ 2 these rapidity
distributions normalized to the rapidity at $y=0$, after integrating
over the $p_t$ distributions.  The data points are from the BRAHMS
Collaboration \cite{Yan08} and the solid curve is $dN/dy$ extracted
from the momentum kick model as given by Eq.\ (\ref{dis2}).  One
observes that the rapidity distributions for $pp$ collisions have the
greatest plateau width in $y$, while the early parton rapidity
distribution has a slightly narrower plateau width and is in between
those of the $pp$ and Au+Au central collisions.  This is consistent
with the evolution of the partons from a $pp$-like distribution to the
early parton distribution, and eventually to the inclusive
nucleus-nucleus rapidity distribution that is closer to a Gaussian
shape than a plateau shape.  The comparison indicates that the
rapidity distribution extracted here during the early moments of
jet-medium interactions is at an intermediate stage in the dynamical
evolution process.  Such a viewpoint is further supported by the
observation that the $p_t$ slope parameter of the early partons (the
ridge particles) is in between those of the jet and the inclusive
central Au+Au distributions \cite{Put07}.

\section{ The Centrality Dependence of the Ridge Yield}

The momentum kick model also provides information on the number of
attenuated kicked partons, $f_R \langle N_k \rangle$, for the most
central collisions. It is of interest to examine whether these number
of kick partons in the most central collision is consistent with the
centrality dependence of the ridge yield.  One can follow the
trajectory of the jet, using the extracted number of kicked partons as
a normalization for the most central collision and infer the ridge
yield as a function of the centrality, to investigate whether the
momentum kick model can also describe the ridge yield at other
centralities.

As the centrality dependence of the ridge yield has not been
investigated in connection with the PHENIX ridge data of Ref.\
\cite{Ada08}, we shall use the STAR centrality data \cite{Put07} to
discuss the centrality dependence.  We review here the description of
the centrality dependence in the momentum kick model.  We wish to show
here that the extracted number of partons kicked by the jet is also
consistent with other related phenomenon.

The momentum kick model separates the ridge yield into a geometrical
factor part that depends on the average number of kicked partons
$2f_R\langle N_k\rangle/3$ and another factor of differential
distribution $E dF/d{\bf p}$ in Eq.\ (\ref{eq2}).  The quantity
$\langle N_k \rangle$ depends on the centrality.  

We consider a jet source point at ${\bf b}_0$, from which a
mid-rapidity jet parton originates.  The number of jet-(medium parton)
collisions along the jet trajectory, which makes an angle $\phi_s$
with respect to the reaction plane, is \cite{Won08a}
\begin{eqnarray}
\label{nk}
N_k({\bf b}_0,\phi_s) = \int_0^{\infty} \sigma \, dl
\frac{dN_{\rm parton}}{dV} \left ({\bf b}' ({\bf b}_0,\phi_s) \right ),
\end{eqnarray} 
where 0$<$$l$$<$$\infty$ parametrizes the jet trajectory, $\sigma$ is the
jet-(medium parton) scattering cross section, and $dN_{\rm parton}({\bf
  b}')/{dV}$ is the parton density of the medium at ${\bf b}'$ along
the trajectory $l$.

Jet-(medium parton) collisions take place along different parts of the
trajectory at different $l$ and involve the medium at different stages
of the expansion.  They depend on the space-time dynamics of the jet
and the medium. Assuming hydrodynamical expansion of the fluid in both
the longitudinal and transverse directions and focusing our attention
on mid-rapidity, we can determined the distribution of the number of
jet-(medium parton) collisions $P(N)$ as a function of the transverse
jet source point coordinate ${\bf b}_0$ and the azimuthal angle
$\phi_s$ \cite{Won08a}.  We need to weight the number of kicked medium
particles by the local binary collision number element $d{\bf b}_0
\times dN_{\rm bin}/d{\bf b}_0$.  The normalized probability
distribution $P(N,\phi_s)$ with respect to the number of ridge
particles (or jet-(medium parton) collisions) is
\begin{eqnarray}
\label{Pn}
P \left( N,\phi_s \right ) 
=\frac{1}{N_{\rm bin}} \int d {\bf b}_0 \frac{dN_{\rm bin}}{d {\bf b}_0}
( {\bf b}_0) \delta \left ( N-N_k({\bf b}_0,\phi_s) \right ).
\end{eqnarray}
Thus, the number of ridge particle yield per trigger particle (or the
number of jet-(medium parton) collisions per trigger) at an azimuthal
angle $\phi_s$, averaged over all source points of binary collisions
at all ${\bf b}_0$ points, is \cite{Won08a}
\begin{eqnarray}
\label{eq44}
\bar N_{k} (\phi_s) = \int N P (N, \phi_s) 
e^{-\zeta N}
~dN \biggr / \int P (N, \phi_s) 
e^{-\zeta N}
~dN,
\end{eqnarray}
where $\zeta$ is the exponential index in the ratio of the
fragmentation function after $N$ jet-(medium parton) collisions relative to
the fragmentation function before any collision,
\begin{eqnarray}
e^{-\zeta N}=\frac{D(p^{\rm trig},{\bf p}_j - \sum_n^N {\bf q}_n -{\bf \Delta}_r)}
                 {D(p^{\rm trig},{\bf p}_j)}.
\end{eqnarray}
where ${\bf q}_n$ is the momentum loss at the $n$th jet-(medium parton)
collision and ${\bf \Delta}_r$ is the momentum loss owing to gluon
radiation.  
From these equations, we get the ridge yield $\bar N_{k} (\phi_s)$ per
trigger as
\begin{eqnarray}
\label{nks}
\bar N_{k} (\phi_s) =
\frac{1}{N_{\rm bin}}
\int  d{\bf b}_0
N_k ({\bf b}_0,\phi_s) 
e^{-\zeta N_k }
\frac{dN_{\rm bin}}{d{\bf b}_0}
\biggr /
\frac{1}{N_{\rm bin}}
\int  d{\bf b}_0
e^{-\zeta N_k }
\frac{dN_{\rm bin}}{d{\bf b}_0}.
\end{eqnarray} 
We get the jet quenching measure \cite{Won08a}
\begin{eqnarray}
\label{nzeta}
R_{AA} (\phi_s) =\frac{N_{\rm trig}} {N_{\rm bin}}
=\int P (N, \phi_s) 
e^{-\zeta N}~dN=
\sum_{N\!=\!0}^{N_{\rm max}}\!P(N,\phi_s) e^{-\zeta N},
\end{eqnarray} 
which can also be obtained as
\begin{eqnarray}
R_{AA} (\phi_s) 
= \frac{1}{N_{\rm bin}}
\int
d{\bf b}_0
\exp \{-\zeta~ N_k ({\bf b}_0,\phi_s) \} 
\frac{dN_{\rm bin}}{d{\bf b}_0}.
\end{eqnarray}
After $\bar N_{k} (\phi_s)$ and $R_{AA}(\phi_s)$ have been evaluated,
we can average over all azimuthal angles $\phi_s$ and obtain the ridge
particles (or jet-(medium parton) collisions) per trigger
\begin{eqnarray}
\langle N_{k} \rangle
=\int_0^{\pi/2} d\phi_s  \bar N_{k} (\phi_s)/(\pi/2),
\end{eqnarray}
and
\begin{eqnarray}
\langle R_{ AA} \rangle
=\int_0^{\pi/2} d\phi_s   R_{AA} (\phi_s)/(\pi/2),
\end{eqnarray}
which is usually expressed just as $R_{AA}$.

\begin{figure} [h]
\includegraphics[angle=0,scale=0.60]{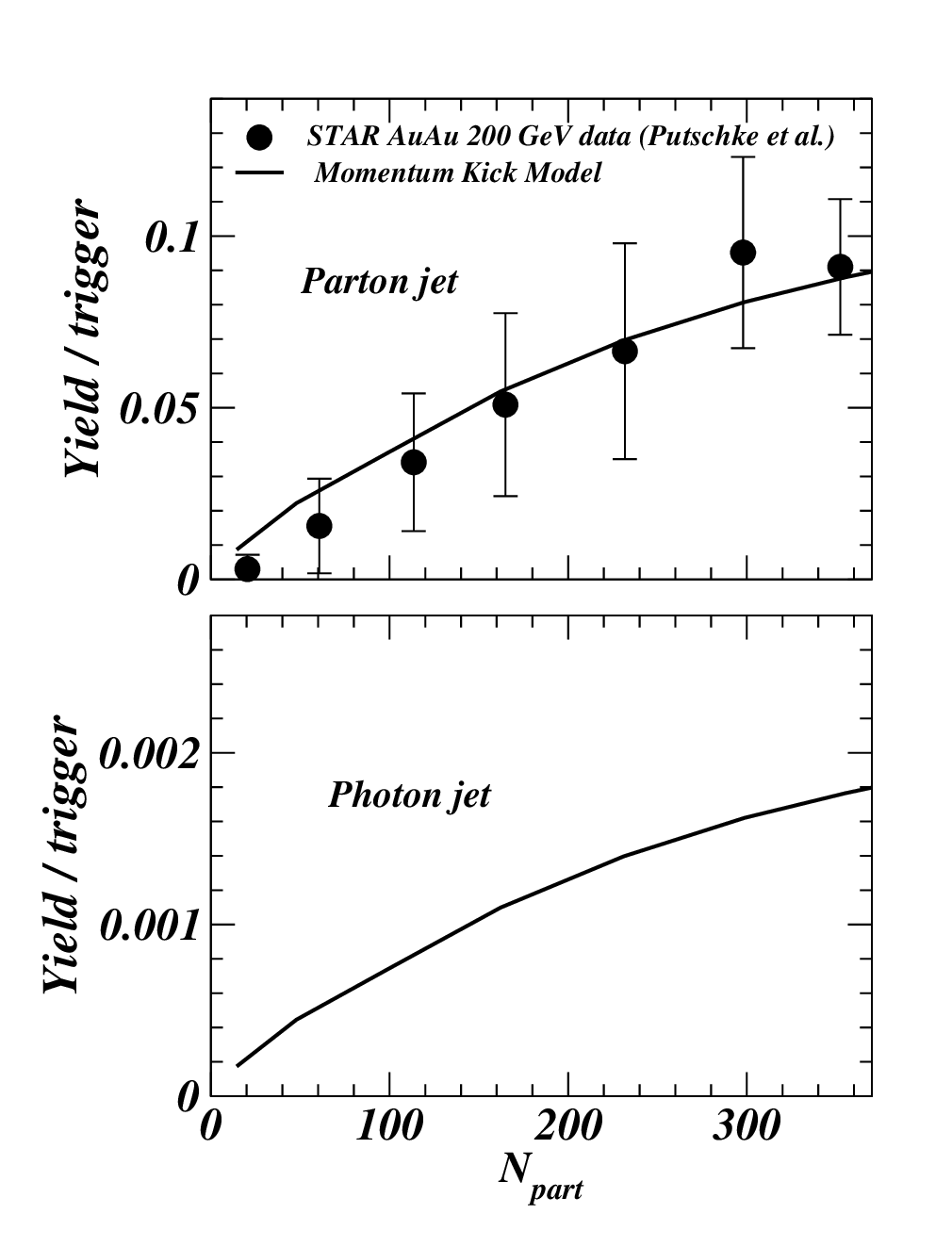}
\vspace*{0.0cm}
\caption{The total yield of charged ridge particles per parton
(hadron) jet (a), and per photon jet (b), as a function of the number of
participants. In Fig.\ 3(a), the data points are from \cite{Put07}. }
\end{figure}

In our previous analysis of the STAR ridge yield and jet quenching, we
find that assuming $f_R=f_J=0.632$, the experimental data of the
centrality dependence of $R_{ AA}$ and the centrality dependence of
the ridge yield using hadron trigger can be explained well when we use
\cite{Won08a}
\begin{eqnarray}
\label{zetapar}
\zeta=0.20, {\rm ~~and~~} \sigma =1.4 {\rm ~~mb}.
\end{eqnarray}
The STAR data of ridge yield per trigger as a function of the number
of participants are shown in Fig. 3(a) and are compared with the
momentum kick model results \cite{Won08a}, for the acceptance of the
STAR Collaboration in \cite{Put07}.  Fig. 3(a) shows that the number
of kicked partons extracted from the most central collision lead us to
a consistent description of the centrality dependence of the ridge
yield, an attenuation index $\xi$ that is consistent with the
fragmentation and energy loss \cite{Won08a}, and a cross section of
1.4 mb.  Note that such a cross section is slightly smaller, but is of
the same order of magnitude, as $\pi a^2$ of the correlation length
extracted from the momentum kick model in Eq.\ (\ref{cor}).  As the
total elastic scattering cross section is a product of the geometrical
cross section and the strength of the potential inside the correlated
region [Eq.\ (116) of Ref.\ \cite{Gla59}], the difference of the cross
section in Eq.\ (\ref{zetapar}) and $\pi a^2$ may provide information
on the depth of the non-perturbative potential relative to which the
parton scatters.

\section{Ridge Particle Yield  from a Photon Jet}

One can consider experiments with two transverse jets in which one of
the two jets is a photon jet on the near side while the other jet is a
strongly interacting parton on the away side.  The use of a near-side
photon jet allows one to probe the origin of the ridge particles as we
discussed in Section I \cite{Cia08}.  If the ridge arises from the
medium as a result of the collision of the near-side jet, as in the
momentum kick model, the substitution of a photon jet for a hadron jet
will lead to a greatly-reduced yield of the ridge particles.  On the
other hand, if the ridge particles arise from ``several extra
particles deposited by forward-backward beam jets into the fireball''
\cite{Shu07} or from the back splash model \cite{Pan07}, then the
ridge particles yield will not be significantly reduced.

We can make a quantitative estimate of the ridge yield in the momentum
kick model for a photon jet that arises from hard-scattering.  The
number of ridge particles depends on the jet-(medium parton) cross
section and the attenuation index $\zeta$.  For the high-$p_t$
photon jet, the photon jet-(medium parton) cross section is
\begin{eqnarray}
\sigma({\rm photon-parton}) = \left (
\frac{\alpha_e}{\alpha_s} \right )^2 \sigma(
  {\rm parton-parton}),
\end{eqnarray}                
where $\alpha_e=1/137$ is the fine-structure constant.  We can take
$\alpha_s=0.2$ as the strong-interaction coupling constant.  With
$\sigma({\rm parton-parton})\sim 1.4$ mb as given
by Eq.\ (\ref{zetapar}), we can estimate
\begin{eqnarray}
\label{phox}
\sigma({\rm photon-parton}) =1.86 \mu b.
\end{eqnarray}                
As the average number of collisions is much less than 1, we can take
$\zeta=0$ in Eq.\ (\ref{nks}) without much error.  One finds that the
ridge yield per photon trigger is then
\begin{eqnarray}
\langle N_k \rangle =
\frac{1}{N_{\rm bin}}
\int \frac{d\phi_s}{(\pi/2)}
d{\bf b}_0
N_k ({\bf b}_0,\phi_s) 
\frac{dN_{\rm bin}}{d{\bf b}_0}.
\end{eqnarray} 
We can evaluate $N_k ({\bf b}_0,\phi_s)$ by using Eq.\ (\ref{nk}) and
the photon-(medium parton) cross section of Eq.\ (\ref{phox}) and obtain
the total number of ridge particle yield per photon jet as a function
of the participant number shown in Fig. 3(b), for the acceptance region
as in \cite{Put07}. The yield for the photon jet is about 0.002 per
photon jet for the most central Au+Au collision, which is small
indeed.  For all practical purposes, a high-$p_t$ photon jet does not
lead to significant production of ridge particles in the momentum kick
model.

\section{Conclusion and Discussions}

Using the momentum kick model, we examine the PHENIX near-side ridge
particle data for central Au+Au collisions at $\sqrt{s_{NN}}=200$ GeV
which  cover the range of pseudorapidity, $|\eta| < 0.35$,
and a large number of $p_t^{\rm trig}\otimes p_t^{\rm assoc}$ combinations.
We find that the PHENIX data can be described by the momentum kick
model.

With the successful analysis of the ridge data from STAR, PHOBOS, and
PHENIX Collaborations, it is of great interest to find out whether the
extracted physical quantities are compatible with those in other
relevant physical phenomena.  The most important quantity extracted is
the (average) magnitude of the longitudinal momentum kick $q_L$ along
the jet direction imparted on the medium parton by the jet parton in a
parton-parton collision.  We find that such a quantity is related to
the momentum transfer squared $t$ of the incident jet parton.  The
magnitude of $|t|$ is less than 1 GeV$^2$, indicating that the
scattering is within the realm of non-perturbative QCD.  The
scattering of the jet parton and a medium parton is characterized by a
correlation length of $a=$0.39 fm.

On the theoretical side, the correlation length $a$ in parton-parton
scattering has been previously obtained in many previous analyses of
hadron-hadron elastic differential cross sections and the model the
non-perturbative pomeron.  The modified Chou-Yang model \cite{Sch81},
the model of the non-perturbative pomeron in terms of the Stochastic
Vacuum \cite{Ber99}, and lattice gauge calculations of the gluon
condensate correlator in \cite{Meg99,Ber99} give a correlation lengths
in the range 0.25-0.37 fm, compatible with the magnitude of the
correlation length extracted in the momentum kick model.  It is
reasonable to conclude that the parton-parton scattering between the
jet parton and the medium parton arises from the exchange of a
non-perturbative pomeron.

It should be emphasized that our ability to ascertain the
non-perturbative nature of the parton-parton scattering is important
in helping us select the proper description for the dynamics of the
interaction of the jet and the medium.  For jet partons in the
momentum range of $p_t^{\rm jet} < 10$ GeV as considered in
measurements involving associated particles, a plausible description
needs to include the non-perturbative aspects of the scattering
between the jet parton and the medium parton, if one wishes to
describe the jet momentum loss and the scattered medium partons
properly.

We can examine further the shape of the rapidity plateau in the early
parton momentum distribution obtained here.  The presence of a
rapidity plateau in early history of a central nucleus-nucleus
collision as inferred from the momentum kick model is not a surprising
result, as the rapidity plateau structure occurs in elementary process
involving the fragmentation of flux tubes \cite{
  Cas74,Bjo83,Won91,Won94,And83} and in many particle production
models
\cite{Wol80,Van88,Odo80,And83,Cap78,Wer88,Sor89,Won89,Wan94,Gei92,McL94,Li96,Jeo97}.
Experimental evidence for a plateau in rapidity distributions has been
observed earlier in $\pi^\pm$ production in high-energy $e^+$-$e^-$
annihilation \cite{Aih88,Hof88,Pet88,Abe99,Abr99} and in $pp$
collisions at RHIC energies \cite{Yan08}.  A comparison of the plateau
structure of $pp$, central Au+Au, and early parton distributions in
Fig. 2 places rapidity distribution extracted here as an intermediate
stage of the dynamical evolution process, just as indicated by the
intermediate value of the inverse slope of the ridge particles between
those of the jet and the inclusive particles.

The number of kicked partons extracted here also provide the proper
normalization to explore the centrality dependence of the ridge yield,
whereas the attenuation index $\xi$ is compatible with the estimates
from fragmentation process of the jet parton.

It is of interest to propose the use of high-$p_t$ photon jets to
examine the associated particles.  In the momentum kick model, the
collision of a high-$p_t$ hadron jet with the medium partons lead to
the recoil of the medium partons which subsequently materialize as
ridge particles.  However, for a high-$p_t$ photon jet the
photon-(medium parton) cross section is greatly reduced, leading to a
much smaller number of produced ridge particles.  Thus, a photon jet
on the near-side will lead to a very small yield of ridge particles.
Such a feature may be used to discriminate among different models.

In summary, we have analyzed PHENIX near-side ridge data for central
Au+Au collisions at $\sqrt{s_{NN}}=200$ GeV.  We found that the data
can be described well by the momentum kick model and the extracted
physical quantities provide useful information on the nucleus-nucleus
collision process.  Specifically, the scattering between the jet
parton and the medium parton arises from the exchange of a
non-perturbative pomeron for $p_t^{\rm jet}< 10$ GeV.  This however is
only the first two step in the theoretical analysis.  The final third
step consists of the construction of theoretical models that can
explain these physical quantities.  Another step is to connect the
observed physical quantities to other observables such as the momentum
distribution of the bulk matter at subsequent stages of the
nucleus-nucleus collision.  The momentum kick model can be further
improved with additional inclusion of other effects such as the
collective flow, a better description of the elementary jet-(medium
parton) collision processes, and perhaps a better Monte Carlo tracking
of the jet trajectory and kicked partons.

\vspace*{0.3cm} The author wishes to thank Drs.\ Vince Cianciolo,
Fuqiang Wang, Jiangyong Jia, and Chin-hao Chen for helpful discussions
and communications.  This research was supported in part by the
Division of Nuclear Physics, U.S. Department of Energy, under Contract
No.  DE-AC05-00OR22725, managed by UT-Battelle, LLC.


\vspace*{-0.3cm}

\begin{thebibliography}{99}
\bibitem{Ada05}
J. Adams $et~al.$ for the STAR Collaboration, 
Phys. Rev. Lett. {\bf 95}, 152301 (2005). 

\bibitem{Ada06}
J. Adams $et~al.$ (STAR Collaboration), 
Phys. Rev. C {\bf 73}, 064907 (2006).

\bibitem{Put07} J. Putschke (STAR Collaboration), 
J. Phys. {\bf G34}, S679 (2007).

\bibitem{Bie07} J. Bielcikova (STAR Collaboration), 
J. Phys. {\bf G34}, S929 (2007).

\bibitem{Wan07} 
F. Wang (STAR Collaboration), Invited talk at the XIth
International Workshop on Correlation and Fluctuation in Multiparticle
Production, Hangzhou, China, November 2007, [arXiv:0707.0815].

\bibitem{Bie07a}.
 J. Bielcikova (STAR Collaboration), Phys.G34:S929-930,2007;
 J. Bielcikova for the STAR Collaboration, Talk
presented at 23rd Winter Workshop on Nuclear Dynamics, Big Sky,
Montana, USA, February 11-18, 2007, [arXiv:0707.3100];
J. Bielcikova for the STAR Collaboration, Talk presented at XLIII
Rencontres de Moriond, QCD and High Energy Interactions, La Thuile,
March 8-15, 2008, [arXiv:0806.2261].

\bibitem{Abe07} B. Abelev (STAR Collaboration), Talk presented at 23rd
Winter Workshop on Nuclear Dynamics, Big Sky, Montana, USA, February
11-18, 2007, [arXiv:0705.3371].

\bibitem{Mol07}
L. Molnar (STAR Collaboration), J. Phys. G {\bf 34}, S593 (2007).

\bibitem{Lon07} R. S. Longacre (STAR Collaboration),
  Int. J. Mod. Phys. E{\bf 16}, 2149 (2007).


\bibitem{Nat08} C. Nattrass (STAR Collaboration), 
J. Phys. G {\bf 35}, 104110  (2008).  

\bibitem{Fen08}
A. Feng, (STAR Collaboration), 
J. Phys. G {\bf 35}, 104082  (2008).

\bibitem{Net08} P. K. Netrakanti (STAR Collaboration)
J. Phys. G {\bf 35}, 104010  (2008).

\bibitem{Bar08}
O. Barannikova (STAR Collaboration), 
J. Phys. G {\bf 35}, 104086  (2008).

\bibitem{Ada08}
A. Adare, $et~al.$ (PHENIX Collaboration), 
Phys. Rev. C {\bf 78}, 014901 (2008).

\bibitem{Mcc08}
M. P. McCumber  (PHENIX Collaboration), 
J. Phys. G {\bf 35}, 104081 (2008).

\bibitem{Che08} Chin-Hao Chen (PHENIX Collaboration), 
``Studying the Medium Response by Two Particle Correlations", Hard Probes 2008
  Intern. Conf. on Hard Probes of High Energy Nuclear Collisions, A
  Toxa, Galicia, Spain, June 8-14, 2008.

\bibitem{Wen08} E. Wenger (PHOBOS Collaboration), J. Phys. G {\bf 35},
  104080 (2008).

\bibitem{Tan08}
M.J. Tannenbaum, 
Eur. Phys. J. C {\bf 61}, 747 (2009).

\bibitem{Jia08qm}
Jiangyong Jia, (PHENIX Collaboration), 
J. Phys. G {\bf 35}, 104033 (2008).

\bibitem{Lee08}
M. van Leeuwen, (STAR Collaboration), 
Eur. Phys. J. C {\bf 61}, 569 (2009).

\bibitem{Dau08}
M. Daugherity, 
(STAR Collaboration), 
J. Phys. G {\bf 35}, 104090  (2008).

\bibitem{Won07}
C. Y. Wong, Phys. Rev. C {\bf 76}, 054908  (2007).

\bibitem{Won07a} C. Y. Wong, Chin. Phys. Lett. {\bf 25}, 3936 (2008)

\bibitem{Won08} C. Y. Wong, J. Phys. G {\bf 35}, 104085 (2008).

\bibitem{Won08a}
C. Y. Wong, Phys. Rev. C {\bf 78}, 064905 (2008).

\bibitem{Won09}
C. Y. Wong, arXiv:0903.3879.

\bibitem{Shu07} E. Shuryak, Phys. Rec. C {\bf 76}, 047901 (2007).

\bibitem{Vol05}
S. A. Voloshin, Nucl. Phys. {\bf A749},  287  (2005).

\bibitem{Chi08}
C. B. Chiu and R. C. Hwa
Phys. Rev. C {\bf 79},  034901 (2009).

\bibitem{Hwa03}
R. C. Hwa and C. B. Yang, Phys.Rev. C {\bf 67} 034902  (2003); 
R. C. Hwa and Z. G. Tan, Phys. Rev. C {\bf 72},  057902 (2005);
R. C. Hwa and C. B. Yang, [nucl-th/0602024].

\bibitem{Chi05}
C. B. Chiu and R. C. Hwa
Phys. Rev. C {\bf 72},  034903 (2005).


\bibitem{Hwa07}
R. C. Hwa, [arXiv:0708.1508].

\bibitem{Pan07} 
V. S. Pantuev, [arXiv:0710.1882].

\bibitem{Dum08}
A. Dumitru, F. Gelis, L. McLerran, and R. Venugopalan,
Nucl. Phys. A{\bf 810}, 91 (2008). 

\bibitem{Gav08}
S. Gavin, and G. Moschelli, J. Phys. G {\bf 35}, 104084 (2008).

\bibitem{Gav08a}
S. Gavin, L. McLerran, and G. Moschelli, Phys. Rev. C {\bf 79}, 051902 (2009).

\bibitem{Arm04} N. Armesto, C. A. Salgado, U. A. Wiedemann,
Phys. Rev. Lett. {\bf 93}, 242301 (2004).

\bibitem{Rom07}
P. Romatschke, Phys. Rev. C {\bf 75} 014901  (2007).

\bibitem{Maj07} A. Majumder, B. Muller, and S. A. Bass,
  Phys. Rev. Lett. {\bf 99}, 042301 (2007).

\bibitem{Dum07} A. Dumitru, Y. Nara, B. Schenke, and M. Strickland,
  Phys. Rev. C {\bf 78}, 024909 (2008); B. Schenke, A. Dumitru,
  Y. Nara, M. Strickland, J. Phys. G {\bf 35}, 104109 (2008).

\bibitem{Miz08}
R. Mizukawa, T. Hirano, M. Isse, Y. Nara, and A. Ohnishi,
J. Phys. G {\bf 35}, 104083 (2008).

\bibitem{Jia08} Jianyong Jia and R.. Lacey, 
 Phys. Rev. C {\bf 79}, 011901 (2009).

\bibitem{Jia08a} Jianyong Jia, 
Eur. Phys. J. C {\bf 61}, 255 (2009).

\bibitem{Kac05} O. Kaczmarek and F.  Zantow, 
Phys. Rev. D {\bf 71}, 114510 (2005).

\bibitem{Won02} 
C. Y. Wong, Phys. Rev. C {\bf 65},  034902 (2002); 
C. Y. Wong, Phys. Rev. C {\bf 65},  014903 (2002);
C. Y. Wong, J. Phys.  G {\bf 28}, 2349  (2002);
C. Y. Wong, Phys. Rev. C {\bf 72},  034906 (2005). 
C. Y. Wong, Talk presented at Recontre de Blois, Chateau de Blois,
France, May 15-20, 2005 [arXiv:hep-ph/0509088];
C. Y. Wong, Phys. Rev. C {\bf 76}, 014902 (2007); 
C. Y. Wong, J. Phys. G {\bf 32},  S301 (2006).

\bibitem{Sch81}
A. Schiz $et~al.$, Phys. Rev. D{\bf 24}, 26 (1981).

\bibitem{Cho68}
T. T. Chou and C. N. Yang, Phys. Rev. {\bf 170}, 1591 (1968). 

\bibitem{Bia76}
A. Bialas $et~al.$, Acta Phys. Pol. {\bf B8},
855 (1977). 

\bibitem{Lev74} E. M. Levin and Shekhter, in Proceedings of the IXth
 Winter LNPI School on Nuclear Physics and Elementary Particles,
 Leningrad, 1974.


\bibitem{Lan88}
P. V. Landshoff and O. Nachtmann,
Zet. Phys. {\bf C35}, 405 (1988).

\bibitem{Nac91}
O. Nachtmann,
Ann. Phys. (N.Y.) {\bf 209}, 436 (1991).

\bibitem{Dos87} H.G. Dosch, Phys. Lett. {\bf B190}, 177 (1987),
 H.G. Dosch, Yu.A. Simonov, Phys. Lett. {\bf B205}, 339 (1988),
 Yu.A. Simonov, Nucl. Phys. {\bf B307}, 512 (1988).

\bibitem{Kra90} A. Kr\"amer and H.G. Dosch, Phys. Lett.B{\bf 252},
  669 (1990).

\bibitem{Dos92}
 H.G. Dosch, E. Ferreira, and A. Kr\" amer, Phsy. Rev. {\bf D50},
 1992 (1992).

\bibitem{Dos01} H. G. Dosch, O. Nachtmann, T. Paulus, and
S. Weinstock, Eur. Phys. J {\bf C21}, 339 (2001).

\bibitem{Ber99}
E. R. Berger and O. Nachtmann,
Eur. Phys. J  {\bf C7}, 459 (1999).  
! a=0.33 - 0.37 fm

\bibitem{DiG92} A. DiGiacomo, H. Panagopoulos, Phys. Lett. {\bf B285},
133 (1992); A. DiGiacomo, H. Panagopoulos, and E. Meggiolaro,
Phys. Lett. {\bf B285}, 133 (1992).

\bibitem{Meg99} E. Meggiolaro, Phys. Lett. {\bf B451}, 414 (1999).

\bibitem{For97}
J.R. Forshaw and D. A. Ross,
{\it Quantum Chromodynamics and the Pomeron}, Cambridge University Press, 1997.

\bibitem{Don02}
S. Donnachie, G. Dosch, P. V. Landshoff and O. Nachtmann,
{\it Pomeron and QCD}, Cambridge University Press, 2002.

\bibitem{Cas74} A. Casher, J. Kogut, and L. Susskind, Phys. Rev. D
{\bf 10}, 732 (1974).

\bibitem{Bjo83}
J. D. Bjorken, Phys. Rev. D {\bf 27}, 140 (1983).

\bibitem{Won91} C. Y. Wong, R. C. Wang, and C. C. Shih,
Phys. Rev. D {\bf 44}, 257 (1991).

\bibitem{Won94} C. Y. Wong, {\it Introduction to High-Energy Heavy-Ion
Collisions}, World Scientific Publishing Company, 1994.


\bibitem{Wol80}
S. Wolfram, Proc. 15th Recontre de Moriond (1980), ed. by Tran Thanh Van;
G. C. Fox and S. Wolfram,  B {\bf 168}, 285 (1980); B. R. Webber,
Nucl. Phys. {\bf 238}, 492 (1984).

\bibitem{Van88}
L. Van Hove,  A. Giovannini,
Acta Phys. Polon. {\bf B19}, 931 (1988);
M. Garetto, A. Giovannini, T. Sjostrand,
L. van Hove, CERN Report CERN-TH-5252/88,
Presented at Perugia Workshop on Multiparticle Dynamics, Perugia,
Italy, Jun 21-28, 1988;  Y. L. Dokshizer, V. A. Khoze, S. I. Troyan,
in {\sl Perturbative Quantum Chromodynamics}, Ed. by A. H. Mueller, World
Publishing, Singapore 1989, p. 241.


\bibitem{Odo80} R. O'dorico, Nucl. Phys. {\bf B172}, 157 (1980);
R. O'dorico, Comp. Phys. Comm.

\bibitem{And83} B. Andersson, G. Gustafson, and T. Sj\"ostrand,
  Zeit. f{\"u}r Phys. {\bf C20}, 317 (1983); B. Andersson,
  G. Gustafson, G.  Ingelman, and T.  Sj\"ostrand, Phys.  Rep. {\bf
    97}, 31 (1983); T. Sj\"ostrand and M. Bengtsson, Computer Physics
  Comm.  {\bf 43}, 367 (1987); B. Andersson, G. Gustavson, and
  B. Nilsson-Alqvist, Nucl. Phys. {\bf B281}, 289 (1987).

\bibitem{Cap78}
A. Capella and A. Krzywicki, Phys. Rev. {\bf D18}, 3357 (1978);
A. Capella and J. Tran Thanh Van, 
Zeit. f{\"u}r Phys. {\bf C10}, 249 (1981);
A. Capella $et~ al.$, 
Zeit. f{\"u}r Phys. {\bf C33}, 541 (1987);
A. Capella
  and J. Tran Thanh Van, Zeit. Phys. {\bf C38}, 177 (1988); 
N. Armesto and C. Pajares, Int. J. Mod. Physi {\bf A15},
  2019 (2000). 

\bibitem{Wer88}
 K. Werner,  Phys. Rev. {\bf D39}, 780 (1988).

\bibitem{Sor89}
H. Sorge, H. St\" ocker, and
  W. Greiner, Nucl. Phys. {\bf A498}, 567c (1989); 

\bibitem{Won89} C. Y. Wong, and Z. D. Lu,
  Phsy. Rev. {\bf D39}, 2606 (1989). 


\bibitem{Wan94}
X. N. Wang and M. Gyulassy, Comp. Phys. Comm. {\bf 83}, 307
  (1994).

\bibitem{Gei92}
K. Geiger and  B. M\"uller, Nucl. Phys. {\bf A544}, 467c (1992);

\bibitem{McL94} 
L. D. McLerran and R. Venugopalan, Phys. Rev. {\bf D49}, 2233 (1994); 
L. D. McLerran and R. Venugopalan, Phys. Rev. {\bf D49}, 3352 (1994); 
L. D. McLerran and R. Venugopalan, Phys. Rev. {\bf D50}, 2225 (1994).
A. Kovner, L. McLerran, and H. Weigert, Phsy. Rev. {\bf D52}, 6231 (1995).


\bibitem{Li96}
Zi-Wei Lin, Che Ming Ko, Bao-An Li and Bin Zhang, and Subrata Pal,
Phy. Rev. {\bf C72}, 064901 (2005).

\bibitem{Jeo97}
S. Jeon and J. Kapusta, Phsy. Rev. {\bf C56}, 468 (1997).


\bibitem{Sch62}
J. Schwinger, Phys. Rev. {\bf 128}, 2425 (1962); J. Schwinger, in
$Theoretical$ $Physics$,
Trieste Lectures, 1962 (I.A.E.A., Vienna, 1963), p. 89.

\bibitem{Low71} J. H. Lowenstein and J. A. Swieca, Ann. Phys. (N.Y.) {\bf 68},
172 (1971).

\bibitem{Col75}
S. Coleman, R. Jackiw, and L. Susskind, Ann. Phys. {\bf 93}, 267 (1975).

\bibitem{Col76}
S. Coleman, Ann. Phys. {\bf 101}, 239 (1976).


\bibitem{Aih88} 
H. Aihara $et~al.$ (TPC/Two\_Gamma Collaboration),
Lawrence Berkeley Laboratory Report LBL-23737 (1988).

\bibitem{Hof88}
W. Hofmann, Ann. Rev. Nucl. Sci. {\bf 38}, 279 (1988).

\bibitem{Pet88} A. Petersen $et~al.$ (Mark II Collaboration),
Phys. Rev. D {\bf 37}, 1 (1988).

\bibitem{Abe99} K. Abe $et~al.$ (SLD Collaboration), Phys. Rev. D {\bf
59}, 052001 (1999).

\bibitem{Abr99} K. Abreu $et~al.$ (DELPHI Collaboration), Phys. Lett.
B {\bf 459}, 397 (1999).

\bibitem{Yan08} Hongyan Yang (BRAHMS Collaboration), J. Phys. G{\bf
35}, 104129 (2008); K. Hagel (BRAHMS Collaboration), APS DNP 2008,
Oakland, California, USA Oct 23-27, 2008.

\bibitem{Jetxxx} J. Casalderrey-Solana, and C. A. Salgado,
  Acta. Phys. Polon. {\bf B38}, 3731 (2007); S. Wicks and M. Gyulassy,
  J. Phys. G{\bf 34}, S989 (2007); M. Gyulassy, P. Levai, and
  I. Vitev, Nucl. Phys. {\bf B 594}, 371 (2001); S. Wicks,
  S. Horowitz, W. Djordjevic, and M. Gyulassy, Nucl.  Phys. A {\bf
    784}, 426 (2007); I. Vitev, Phys.Lett. {\bf B639}, 38 (2006);
  E. Wang and X. N. Wang, Phys. Rev. Lett. {\bf 87}, 142301 (2001);
  E. Wang and X. N. Wang, Phys. Rev. Lett. {\bf 89}, 162301 (2002);
  A. Drees, H. Feng, and J. Y.  Jia, Phys. Rev. {\bf C71}, 034909
  (2005); G. Wang and H. Huang, Phys. Lett. B{\bf 672}, 30 (2009).

\bibitem{Cia08} The author wishes to thank Dr. Vince Cianciolo for
  suggesting the use of the near-side photon jet to distinguish
  different ridge models.


\bibitem{Sjo01}
T. Sj\" ostrand $et~al.$, Computer Physics Commun. {\bf 135}, 238 (2001).

\bibitem{Jia09a}
Jiangyong Jia, for the PHENIX Collaboration,
arXiv:0906.3776

\bibitem{Low75} F. E. Low, Phys. Rev. D{\bf 12}, 163 (1975);
S. Nussinov, Phys. Rev. Lett. {\bf 34}, 1286 (1975).


\bibitem{Dol92} J. Dolejsi and J. Hufner, Z. Phys. {\bf C54}, 489
(1992).

\bibitem{Zha03} Wei-Ning Zhang and C. Y. Wong, Phys.Rev. {\bf C68},
035211 (2003).


\bibitem{Shi79} M. A. Shifman, A. I. Vainshtein, and V. I. Akharov,
Nucl. Phys. {bf B147}, 385, 519 (1979).

\bibitem{Nar95}
S. Narison, Nucl. Phys. Pro. Supplement {\bf 54A}, 238 (1997).



\bibitem{Gla59}
R. J. Glauber, ``High-Energy Collision Theory'', in {\it Lectures in
Theoretical Physics}, edited by W. E. Brittin and L. G. Dunham
(Interscience, N.Y., 1959), Vol. 1, p. 315.

\end{thebibliography}
\end{document}